\documentclass[11pt, a4paper]{article}
\pdfoutput=1

\usepackage[utf8]{inputenc}
\usepackage{graphicx}
\usepackage{geometry}
\usepackage[english]{babel}

\usepackage{tikz}
\usetikzlibrary{decorations.pathmorphing}
\usepackage{pgfplots}
\usepgfplotslibrary{patchplots}
\usepackage{subcaption} 

\usepackage[numbers,sort&compress]{natbib}
\usepackage{epsfig}
\usepackage{amsmath}
\usepackage{amssymb}
\usepackage{amsfonts}
\usepackage{anysize}
\usepackage{latexsym}
\usepackage{xcolor}
\usepackage[pdftex,colorlinks=true,linkcolor=blue,citecolor=blue]{hyperref}
\hypersetup{linktocpage}
\usepackage{float}
\usepackage{dsfont}
\numberwithin{equation}{section}

\definecolor{ccqqqq}{rgb}{1,0.5,0}
\definecolor{uuuuuu}{rgb}{0.26666666666666666,0.26666666666666666,0.26666666666666666}
\definecolor{qqwwzz}{rgb}{0,0.3,0.9}

\newcommand{\beq}{\begin{equation}}
\newcommand{\eeq}{\end{equation}}
\newcommand{\bea}{\begin{eqnarray}}
\newcommand{\eea}{\end{eqnarray}}
\newcommand{\bit}{\begin{itemize}}
\newcommand{\eit}{\end{itemize}}
\def\nl{\nonumber \\}

\def\a{\alpha}

\def\b{\beta}

\def\l{\lambda}

\def\p{\partial}

\def\le{\left(}
\def\ri{\right)}

\newcommand{\sn}[2]{\text{sn}\left(\left.\! #1\right| #2\right)}

\newcommand{\cn}[2]{\text{cn}\left(\left.\! #1\right| #2\right)}

\newcommand{\dn}[2]{\text{dn}\left(\left.\! #1\right| #2\right)}

\renewcommand{\a}{\alpha}
\renewcommand{\b}{\beta}

\renewcommand{\d}{\delta}
\renewcommand{\th}{\theta}

\newcommand{\e}{\epsilon}

\renewcommand{\l}{\lambda}

\newcommand{\vf}{\varphi}

\newcommand{\elF}[2]{\mathds{F}\left(\left.#1\right|#2\right)}
\newcommand{\elE}[2]{\mathds{E}\left(\left.#1\right|#2\right)}
\newcommand{\eE}{\mathds{E}}
\newcommand{\eK}{\mathds{K}}

\newcommand{\eP}[2]{\Pi\left(\left.#1\right|#2\right)}
\date{}

\begin{document}

\begin{titlepage}

\begin{flushright}

\end{flushright}
\bigskip
\begin{center}
{\LARGE  {\bf
Action complexity in the presence of defects \\ and boundaries
  \\[2mm] } }
\end{center}
\renewcommand{\thefootnote}{\fnsymbol{footnote}}
\bigskip
\begin{center}
 {\large \bf Roberto Auzzi$^{a,b}$},
 {\large \bf Stefano Baiguera$^{c}$}, \\
  {\large \bf Sara Bonansea$^{d}$}  {\large \bf and} 
  {\large \bf Giuseppe Nardelli$^{a,e}$}
\vskip 0.20cm
\end{center}
\vskip 0.20cm 
\begin{center}
$^a${ \it \small  Dipartimento di Matematica e Fisica,  Universit\`a Cattolica
del Sacro Cuore, \\
Via della Garzetta 48, 25133 Brescia, Italy}
\\ \vskip 0.20cm 
$^b${ \it \small{INFN Sezione di Perugia,  Via A. Pascoli, 06123 Perugia, Italy}}
\\ \vskip 0.20cm 
$^c${ \it \small{Department of Physics, Ben-Gurion University of the Negev, \\ Beer Sheva 84105, Israel}}
\\ \vskip 0.20cm 
$^d${ \it \small{The Niels Bohr Institute, University of Copenhagen, \\
Blegdamsvej 17, DK-2100 Copenhagen \O, Denmark}}
\\ \vskip 0.20cm 
$^e${ \it \small{TIFPA - INFN, c/o Dipartimento di Fisica, Universit\`a di Trento, \\ 38123 Povo (TN), Italy} }
\\ \vskip 0.20cm 
E-mails: roberto.auzzi@unicatt.it, baiguera@post.bgu.ac.il,  \\
sarabonansea@gmail.com, giuseppe.nardelli@unicatt.it
\end{center}
\vspace{3mm}

\begin{abstract}
\begin{itemize}
The holographic complexity of formation for the AdS$_3$ $2$-sided Randall-Sundrum model and the 
AdS$_3$/BCFT$_2$ models is logarithmically divergent according to the volume conjecture, while it is finite using the action proposal.
One might be tempted to conclude that
the UV divergences of the volume and action conjectures are always different for defects and boundaries in two-dimensional conformal field theories.
We show that this is not the case. In fact, in Janus AdS$_3$ we find that both volume and action proposals provide the same
kind of logarithmic divergences.

\end{itemize}
\end{abstract}

\end{titlepage}

\tableofcontents


\section{Introduction}

Starting from the seminal work by Ryu and Takayanagi \cite{Ryu:2006bv} 
on the holographic dual of Entanglement Entropy (EE),
the development of the AdS/CFT correspondence \cite{Maldacena:1997re}
intertwined with quantum information.
One of the characters that recently entered the scene is computational complexity,
which may provide a field theory dual to the asymptotic growth of the Einstein-Rosen Bridge (ERB) 
after long time scales \cite{Susskind:2014rva,Susskind:2014moa}.
Heuristically, quantum computational complexity estimates the difficulty to build a target state starting from a simple, usually unentangled, reference state.
This is done by counting the number of steps needed to reach the target state from the reference one,
 picking unitaries from a universal set of elementary operations \cite{Nielsen1,Nielsen2}.
This problem is of primary importance in the context of quantum infomation \cite{Aaronson:2016vto}.
Two main conjectures have been proposed as  holographic duals of computational complexity:
\begin{itemize}
\item Complexity=volume (CV)  \cite{Stanford:2014jda}, in which 
 complexity is proportional
 to the volume of the maximal slices anchored to the boundary 
\beq
\mathcal{C}_V \sim \frac{\mathcal{V}}{G L} \, ,
\eeq
where $\mathcal{V}$ is the maximal volume of the ERB, $G$ the Newton's constant and $L$ the AdS radius.
\item Complexity=action (CA)   \cite{Brown:2015bva,Brown:2015lvg}, in which
complexity is proportional  to the gravitational action evaluated on
 the Wheeler DeWitt  (WDW) patch, which is the bulk domain of  dependence
 of the above-mentioned spatial slice 
\beq
\mathcal{C}_A = \frac{I_{\rm WDW}}{\pi \hbar} \, ,
\eeq
where $I_{\rm WDW}$ is the on-shell gravitational action evaluated on
 the WDW patch.  We will use natural units for the Planck's constant $\hbar =1.$
\end{itemize}
In spite that both the CV and the CA proposals have been investigated 
in several contexts \cite{Lehner:2016vdi,Carmi:2016wjl,Chapman:2016hwi,Couch:2016exn,Cai:2016xho,Reynolds:2016rvl,
Carmi:2017jqz,Auzzi:2018zdu,Auzzi:2018pbc,Alishahiha:2018tep,Bolognesi:2018ion,Bernamonti:2019zyy,Hashemi:2019aop,Bernamonti:2021jyu,Belin:2021bga}, 
we are still far away from a definitive understanding of complexity
conjectures. One of the most important open problems
is a satisfactory definition of the computational complexity 
on the field theory side. Up to now, most of the developments
have been done in quantum-mechanical systems with a finite number of degrees of freedom
 \cite{Brown:2019whu,Balasubramanian:2019wgd,Auzzi:2020idm,Balasubramanian:2021mxo,Brown:2021rmz} 
 and in free field theories \cite{Jefferson:2017sdb,Chapman:2017rqy,Khan:2018rzm,Doroudiani:2019llj},
but a precise definition in interacting CFTs is still lacking 
(see e.g. \cite{Caputa:2017urj,Caputa:2018kdj,Erdmenger:2020sup,Flory:2020eot,Chagnet:2021uvi,Boruch:2021hqs}
for some progresses in this direction). 
See also \cite{Couch:2021wsm} for the case of Topological Quantum Field Theory.
See \cite{Susskind:2018pmk,Chapman:2021jbh} for reviews.

To achieve further insights, it may be useful to take inspiration from the EE,
for which both the holographic and the field theory side of the duality are under control.
In field theory, the definition of EE requires 
a splitting of the system in two complementary subregions.
In the gravity theory, the entropy is computed as the area delimited by the  Ryu-Tayanagi (RT) surface  \cite{Ryu:2006bv},
which is attached on the boundary of the given subsystem.
It is then natural to conjecture that subsystems play an important role also for  complexity.
Indeed, several definitions have been proposed to generalise the concept of computational
complexity to mixed states and subregions \cite{Agon:2018zso,Caceres:2019pgf}. 
On the holographic side, both the volume and action conjectures have natural extensions 
to the case of subsystems. 
The CV generalisation \cite{Alishahiha:2015rta}
 requires to compute the maximal volume of the codimension-one bulk surface $\mathcal{R}_A$ anchored to a subregion $A$ on the boundary and delimited by its Ryu-Tayanagi (RT) surface
\beq
\mathcal{C}_V (A) \sim \frac{\mathcal{V}(\mathcal{R}_A)}{G L} \, .
\eeq
The CA subregion proposal  \cite{Carmi:2016wjl}
 requires instead to calculate the gravitational action in the intersection between the WDW patch and the
  Entanglement Wedge (EW), which is the bulk domain of dependence of the RT surface:
\beq
\mathcal{C}_A (A) = \frac{I_{\rm WDW \cap EW}}{\pi \hbar} \, .
\eeq
Subregion complexity has then been investigated for several configurations \cite{Abt:2017pmf,Auzzi:2019vyh,Chen:2018mcc,Alishahiha:2018lfv,Auzzi:2019fnp,Auzzi:2019mah,DiGiulio:2021oal,DiGiulio:2021noo},
 including the Banados-Teitelboim-Zanelli (BTZ)  \cite{Banados:1992gq} black hole.

At the qualitative level, the volume and the action conjectures share many important features, such as
the linear growth at late time  \cite{Stanford:2014jda,Brown:2015bva,Brown:2015lvg}, the structure of divergences \cite{Reynolds:2016rvl,Carmi:2016wjl} 
and the switch-back effect \cite{Susskind:2014jwa}.
A certain degree of arbitrariness is expected in defining computational complexity,
due to the choice of the reference state and of the allowed computational gates.
Consequently, CV and CA (and their further generalizations \cite{Couch:2016exn,Belin:2021bga})
 might correspond to different ways to define complexity on the field theory side.
It is then crucial to focus on the examples where CV and CA provide different results.
Systems with  defects may provide such examples \cite{Flory:2017ftd}.
Indeed, this is precisely what happens for the $2$-sided Randall-Sundrum  ($2$-RS) model \cite{Randall:1999vf} in AdS$_3$.
In this case, the contribution to the CV due to the defect contains a logarithmic divergence in the UV regulator, 
while CA is not influenced by the presence of the defect  \cite{Chapman:2018bqj}.
This is true both for the complexity of the total space and for the subregion complexity
of an interval centered around the defect, once the subtraction of the vacuum result is performed (complexity of formation).

Boundaries are related to defects via the folding trick \cite{Bachas:2001vj}, and so we expect a similar
behaviour for their contribution to complexity. 
One can consider also the $1$-sided version of the Randall-Sundrum model,
which is dual to a Boundary  Conformal Field Theory (BCFT).
In the following we shall refer to this case as the AdS$_{d+1}$/BCFT$_d$ model
 \cite{Takayanagi:2011zk,Fujita:2011fp, Nozaki:2012qd}.
Complexity in  AdS$_{d+1}$/BCFT$_d$ model 
was investigated in \cite{Braccia:2019xxi,Sato:2019kik}. For $d=2$, the contribution of the defect 
to CV is again logarithmically divergent, while the contribution to CA is finite.
For $d>2$, instead,  both volume and action give rise to the same type of divergences.
These results were established in \cite{Braccia:2019xxi,Sato:2019kik} for the case of total complexity.
The  behaviour  of the UV divergences should be the same
also for the complexity of a subregion which contain the defect, because (by locality) the UV divergences are expected to 
come from the region nearby the defect.
This was explicitly checked in AdS$_3$/BCFT$_2$   for CV in \cite{Braccia:2019xxi}.
In section \ref{sect-computation_action_BCFT}  we will check this claim also for CA. 

Studying these examples, one is tempted to conclude that, 
for defects and boundaries in 2-dimensional field theories, the UV divergences of CV and CA
are different. It is important to understand if this
 is a general feature of every two-dimensional theory with defects.
In this paper, we show that this is not the case.

To this purpose, we study complexity in Janus AdS$_3$.
This geometry is a dilatonic deformation \cite{Bak:2003jk,Bak:2007jm} of pure AdS$_3$,
which  can be embedded in type IIB supergravity.
 Due to technical reasons related to the regularization of IR-divergences,
 in this background it is natural to directly work with the case of subregions.
 In fact, the length of the subregion provides a natural IR regulator.
In \cite{Auzzi:2021nrj} we  considered the volume conjecture for Janus 
AdS$_3$ and we found that, also in this case, the contribution
to complexity due to the defect  is logarithmically divergent.
We performed the calculation with three different regularizations
(Fefferman-Graham, single and double cutoff regularizations \cite{Estes:2014hka,Bak:2016rpn,Gutperle:2016gfe}) 
and we checked that  the coefficient of the logarithmically divergent  term
is independent of the regularization choice.

 In section \ref{sect-complexity-janus} we will study the subregion  action
complexity for Janus AdS$_3$ and we will find that,
contrarily to what happens in the three dimensional $2$-RS and AdS/BCFT models,
the contribution to complexity due the defect is logarithmically divergent,
as it happens for the volume complexity.
We summarise the results for the contribution of the defect 
to CV and CA in various models in table \ref{tab:results}.

\begin{table}[ht]   
\begin{center}    
\begin{tabular}  {|c|c|c|} \hline  &  $\Delta \mathcal{C}_V (l)$ & $\Delta \mathcal{C}_A (l)$  \\ \hline
\rule{0pt}{4.9ex} $2$-sided Randall-Sundrum  & $ \frac{2}{3} c \,  \eta_{\rm RS}  \,
\log \le    \frac{l}{\,  \delta} \ri
 + \text{finite} $  & $ 0 $  \\
\rule{0pt}{4.9ex} AdS$_3$/BCFT$_2$ & $\frac{2}{3} c \,  \eta_{\rm BCFT}  \,
\log \le    \frac{l}{\,  \delta} \ri
 + \text{finite} $  &  $ \mathrm{finite}   $     \\ 
 \rule{0pt}{4.9ex}  Janus AdS$_3$  & $ \frac{2}{3} c \,  \eta_{\rm JAdS}  \,
\log \le    \frac{l}{\,  \delta} \ri
 + \text{finite}  $ & $ \frac{2 c}{3 \pi^2}  P(\gamma, \tilde{L}/ L) \log \le \frac{l}{ \delta} \ri + \mathrm{finite} $   \\[0.2cm]
\hline
\end{tabular}   
\caption{Behaviour of the contributions of the defect $\Delta \mathcal{C}_V$ and  $\Delta \mathcal{C}_A$
 to the subregion complexity,  for an interval of length $l$ for CV and CA, respectively.
The coefficients of the log divergences $\eta$ are specific of the details of the defect or boundary.
In the case of Janus geometry, $\eta_{\rm JAdS}$ is given in eq. (\ref{eq:eta_gamma}). For the other models, the  
 $\eta$ coefficients can be extracted from \cite{Chapman:2018bqj,Braccia:2019xxi,Sato:2019kik,Auzzi:2021nrj};
their specific values are not essential for the present discussion.
The function $P(\gamma,\tilde{L} / L)$ is given in eq.~\eqref{eq:prefactor_log_divergence}.} 
\label{tab:results}
\end{center}
\end{table}

The first comment comes from reading the table by columns.
While in the volume case the three models have a common logarithmic divergence, for the action 
there are three different behaviours:
\begin{itemize}
\item The action of the $2$-RS model is completely blind to the presence of the defect, 
because  it does not depend on the brane tension.
 Therefore, after subtracting the vacuum part of the action, we find an identically zero $\Delta \mathcal{C}_A $.
\item The action of the AdS/BCFT model is modified by the presence of the end-of-the-world brane because the action
depends on the tension of the brane. After subtracting the vacuum contribution, the divergences cancel and $\Delta \mathcal{C}_A$  is finite.
\item In the Janus geometry, the divergent part of the action 
is modified by the presence of the defect. After subtracting the vacuum part, 
the log divergence  survives in $\Delta \mathcal{C}_A $ and depends on the parameter $\gamma$
of the Janus solution.
\end{itemize}
Another perspective that can be taken is to compare the volume and the action results for each background, \emph{i.e.} we read Table \ref{tab:results} by rows.
In this case we note that:
\begin{itemize}
\item The $2$-RS model distinguishes between volume and action: the former has a logarithmic divergence dependent 
on the tension of the defect, while the latter is identically zero.
\item The AdS/BCFT model distinguishes between volume and action, but in a milder way. 
Only the finite term in the action depends on the brane tension.
Furthermore, in higher dimensions ($d>2$) the same divergences reappear
both in the action and in the volume \cite{Sato:2019kik}.
\item For the Janus AdS$_3$ geometry both the volume and the action
 have a logarithmic divergence dependent on the deformation parameter $\gamma$.
\end{itemize}
The manuscript is organized as follows.
In section \ref{sect-preliminaries} we introduce the Janus AdS$_3$ and the AdS$_3$/BCFT$_2$ geometries, 
we list all the terms entering the gravitational action and we discuss the regularization prescriptions to systematically treat UV divergences.
In sections \ref{sect-complexity-janus} and \ref{sect-computation_action_BCFT} 
we derive the results collected in Table \ref{tab:results} by performing the calculation of subregion action complexity 
for the Janus and the BCFT backgrounds, respectively.
Further open problems are discussed in section \ref{sect-conclusion}.
The appendices contain  technical details.

\section{Preliminaries}
\label{sect-preliminaries}

In this section we introduce the main characters
entering the computation of subsystem complexity.
In section \ref{sect-preliminaries_JanusAdS} we will review the Janus AdS$_3$ geometry, dual to an interface CFT (ICFT) where the coupling constant is different on each side of the interface. 
 In  section \ref{sect-preliminaries_BCFT} we will describe the AdS/BCFT model.
 In the remaining sections, we will discuss the gravitational action in the presence of null boundaries and the regularizations adopted in our calculation.

\subsection{Janus AdS$_3$ geometry}
\label{sect-preliminaries_JanusAdS}

The Janus AdS$_3$ geometry is a solution of type IIB supergravity 
 which preserves the isometry subgroup $\mathrm{SO}(1,2) \times \mathrm{SO}(4)$ 
 of the background geometry $\mathrm{AdS}_3 \times S^3 \times M_4,$ where $M_4$ is a four-dimensional compact manifold \cite{Bak:2007jm}.
Upon dimensional reduction  we obtain Einstein gravity coupled to a dilaton field $\phi$, i.e. 
\beq
I = \frac{1}{16 \pi G} \int d^3 x \sqrt{-g} \, \le R - \p^a \phi \p_a \phi + \frac{2}{L^2} \ri  \, ,
\label{azio-janus}
\eeq
where $L$ the AdS$_3$ radius.
The metric of the Janus solution  reads
 \beq
ds_3^2 = L^2 f(\mu)  \cos^2 \mu \, ds^2_{\mathrm{AdS}_3} \, ,
\label{eq:metric_Janus_mucoord}
\qquad
ds^2_{\mathrm{AdS}_3} = \frac{1}{\cos^2 \mu} \le d\mu^2 + ds^2_{\mathrm{AdS}_2} \ri \, .
\eeq
Unless otherwise specified, the two-dimensional AdS slices will be parametrized using Poincaré coordinates
\beq
ds^2_{\mathrm{AdS}_2} = \frac{1}{z^2} \le dz^2 - dt^2 \ri \, .
\eeq
The profile function $f$ and the dilaton $\phi$ are given by \cite{Freedman:2003ax}
\bea
f (\mu) &=& \frac{\alpha_+^2}{\mathrm{sn}^2 \le \alpha_+ (\mu + \mu_0) | m \ri} \, ,  \nl
\phi(\mu) &=& \phi_0 + \sqrt{2} \log \left[  \mathrm{dn} \le \alpha_+ (\mu + \mu_0) | m \ri - \sqrt{m} \, \mathrm{cn} \le \alpha_+ (\mu + \mu_0) | m \ri \right] \, ,
\label{eq:solutions_f_dilaton_mu_coordinates}
\eea
where 
\beq
\alpha_{\pm}^{2}=\frac{1}{2}\left(1 \pm \sqrt{1-2 \gamma^{2}}\right) \, , \qquad
m= \le \frac{\a_-}{\a_+} \ri^2 \, ,
\qquad
 \mu_0 =\frac{ \mathbb{K} (m) }{\alpha_+ }\, .
 \label{eq:data_elliptic_functions}
\eeq
The conventions on the Jacobi elliptic functions are collected in Appendix \ref{app-elliptics}.
The parameter $\gamma \in [0, \frac{1}{\sqrt{2}}]$ 
specify the details of the dilatonic deformation.
The range of the angular coordinate is $\mu \in [-\mu_0, \mu_0]$ with $\mu_0 \geq \pi/2.$ 
The value $\gamma=0$ corresponds to vacuum AdS space with constant dilaton, 
 $\mu_0 = \pi/2$ and $f(\mu) = \frac{1}{\cos^2 \mu}$.
The case $\gamma= \frac{1}{\sqrt{2}}$ corresponds to an infinite dilaton excursion between the two sides of the Janus interface.


In some cases, it is convenient to change variables from $\mu$ to $y$ in the following way
 \beq
 d \mu= \frac{dy}{ \sqrt{f(y)}} \, ,
\qquad
ds^2 = L^2 \le  f(y) ds^2_{\mathrm{AdS_2}} + dy^2 \ri  \, .
\label{eq:Janus_metric}
\eeq
In this coordinate system,  
 $ y \in [-\infty, \infty] $
 and these two extrema correspond to the two sides of the boundary where the dual interface field theory lives.
In this system the profile functions are
\bea
f(y) &=& \frac{1}{2} \le 1 + \sqrt{1-2\gamma^2} \cosh (2y) \ri \, , \nl
\phi(y) &=& \phi_0 + \frac{1}{\sqrt{2}} \log \le \frac{1+\sqrt{1-2\gamma^2} + \sqrt{2} \gamma \tanh y}{1 + \sqrt{1-2\gamma^2}-\sqrt{2} \gamma \tanh y} \ri \, .
\label{eq:function_f_and_dilaton}
\eea
This geometry admits a dual description in terms of a two-dimensional interface CFT 
where the deformation is produced by a marginal operator $O(x)$ 
with couplings  $J_\pm \int d^2x \, O(x)$ on each side of the boundary, such that
  \beq
  J_{\pm} = \lim_{y \rightarrow \pm \infty} \phi(y) \, .
  \eeq
Since the Janus deformation is associated with an exactly marginal operator, it does not change the central charge of the CFT.

\subsection{AdS$_3$/BCFT$_2$ model}
\label{sect-preliminaries_BCFT}

\begin{figure}
\centering
\includegraphics[scale=0.3]{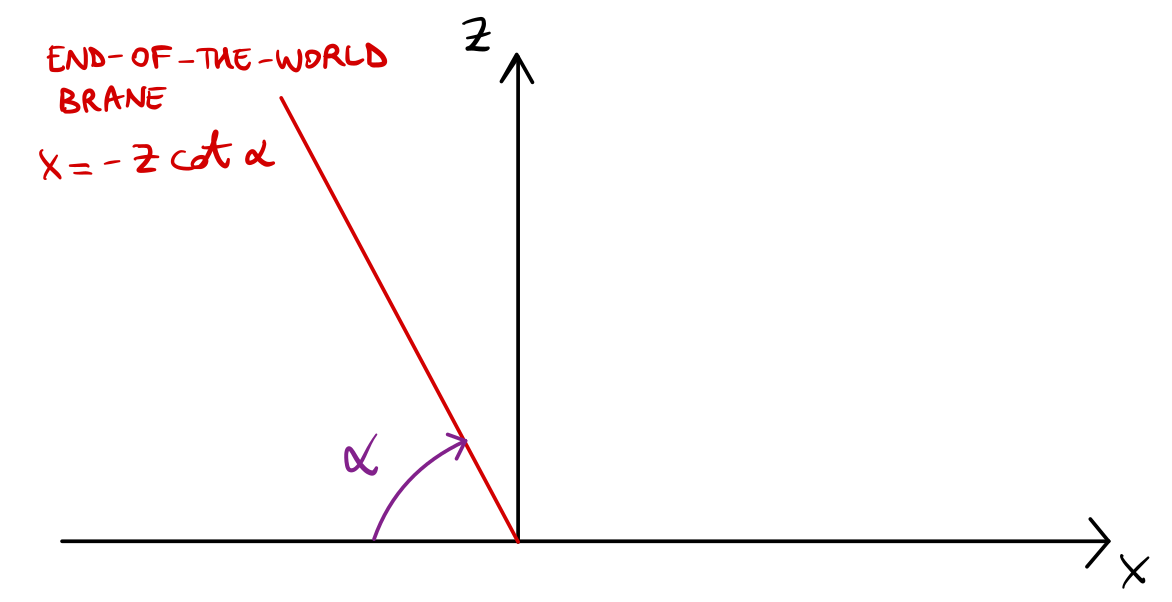}
\caption{Schematic set-up of the AdS/BCFT model. We represent a section at constant time and represent the end-of-the-world brane as the red hypersurface defined by $x=-z \cot \alpha .$}
\label{fig-setup_AdS_BCFT}
\end{figure}

The AdS/BCFT model
can be thought as a $1$-sided version of the Randall-Sundrum setup  \cite{Randall:1999vf},
in which the brane intersects the asymptotically AdS boundary.
The AdS$_3$/BCFT$_2$ model
 was studied in detail in \cite{Takayanagi:2011zk,Fujita:2011fp, Nozaki:2012qd} to describe a QFT
 which is restricted to live on a half plane of flat space, \emph{i.e.} along the portion of spacetime given by $x \geq 0$ in the Minkowski metric, as shown in Fig.~\ref{fig-setup_AdS_BCFT}.
The bulk dual description corresponds to AdS$_{3}$ space with a boundary given by an end-of-the-world brane $\mathcal{Q}$ of tension $T$. 
We use the AdS metric in Poincaré coordinates 
\beq
ds^2 = \frac{L^2}{z^2} \le  -dt^2+  dz^2 + dx^2 \ri \, ,
\label{eq:metric_AdS_BCFT}
\eeq
and the gravitational action is supplemented by a codimension-one term 
\beq
I = \frac{1}{16 \pi G} \int_{\mathcal{B}} d^3 x \, \sqrt{-g} \,
\le R + \frac{2}{L^2} \ri 
+ \frac{1}{8 \pi G} \int_{\mathcal{Q}} d^2 x \, \sqrt{-h} \, \le K - T \ri \, ,
\label{eq:beginning_action_AdS_BCFT}
\eeq
where $\mathcal{B} $ is the bulk AdS spacetime and $\mathcal{Q}$ is the brane located
 at $ x = - z \, \cot \alpha $,
  with induced metric $h_{\mu\nu}$ and trace of the extrinsic curvature $K.$
 The tension of the brane reads
\beq
T= \frac{1}{L} \cos \alpha \, .
\label{eq:tension_brane}
\eeq

\subsection{Gravitational action with null boundaries}

In order to evaluate the gravitational action associated to a subsystem on the boundary, we discuss the contributions coming from null boundaries, following \cite{Lehner:2016vdi}.
The total on-shell action is
\beq
\begin{aligned}
I_{\rm tot} & = \frac{1}{16 \pi G} \int_{\rm WDW \cap EW} d^{d+1} x \, 
\sqrt{-g} \le R+\frac{2}{L^2}   \ri   \\
& + \frac{\varepsilon_{t,s}}{8 \pi G} \int_{\mathcal{B}_{t,s}} 
d^d x \, \sqrt{|h|} \, K 
+ \frac{\varepsilon_n}{8 \pi G} \int_{\mathcal{B}_n} d \lambda \,
 d^{d-1} x \, \sqrt{\gamma} \, \kappa \\
  & + \frac{\varepsilon_{\eta}}{8 \pi G} \int_{\mathcal{J}_{t,s}} d^{d-1} x \, 
 \sqrt{\gamma} \, \eta
 + \frac{\varepsilon_{\mathfrak{a}}}{8 \pi G} \int_{\mathcal{J}_n} d^{d-1} x \, 
 \sqrt{\gamma} \, \mathfrak{a}  \\
 & + \frac{1}{8 \pi G} \int_{\mathcal{B}_n} d\lambda \, d^{d-1} x \,  \sqrt{\gamma} \,
 \Theta \log |\tilde{L} \Theta |  \, ,
\end{aligned}
\label{eq:total_gravitational_action}
\eeq
where $d=2$ in the cases considered in the present work. We comment on each term:
\begin{itemize}
\item The first line contains the bulk term, \emph{i.e.} the Einsten-Hilbert action with cosmological constant, evaluated in the intersection between the WDW patch and the EW.
For the Janus geometry, the bulk term also includes the kinetic part of the dilaton field, see  eq.~(\ref{azio-janus}).
\item The second line contains codimension-one boundary terms which make the variational problem well-defined. 
The first contribution refers to timelike or spacelike surfaces $\mathcal{B}_{t,s}$ and it 
is the Gibbons-Hawking-York (GHY) term, containing the determinant of the induced metric $h$ and the trace $K$ of the extrinsic curvature.
In the case of the AdS/BCFT model, this term is supplemented by a tension term involving the end-of-the-world brane, see eq.~\eqref{eq:beginning_action_AdS_BCFT}.
The second term is evaluated on the null boundaries $\mathcal{B}_n$ and involves
 the integration along the parameter $\lambda$ describing a congruence 
 of null geodesics generating the surface and  the integration along the remaining $(d-1)$ orthogonal directions with induced metric $\gamma.$
The parameter $\kappa$ is defined by the geodesic equation
\beq
k^{\nu} D_{\nu} k^{\mu} = \kappa \, k^{\mu} \, .
\eeq
If the parameter $\lambda$ is affine, $\kappa$ identically vanishes.
The prefactors $\varepsilon_{t,s}$ and $\varepsilon_n$ (referring to timelike/spacelike and null surfaces, respectively) 
take the values $\pm 1$ depending on the orientation of the normals to the hypersurfaces $\mathcal{B}_{t,s}$ or $\mathcal{B}_n$ of interest.
\item The third line contains joint terms, which are codimension-two surfaces found at the intersection of the previous codimension-one boundary terms.
When there is at least a timelike or spacelike surface, they involve the boost parameter 
$\eta,$ while in the purely null case they contain a scalar product of the corresponding null normals, here denoted with $\mathfrak{a}.$ 
We will be more explicit about the expressions of the integrands when doing the actual computations of this work.
\item  The last line is a counterterm which must be included on null boundaries
to restore reparametrization invariance, which is broken by the terms
in the second and third lines. This introduces an extra scale $\tilde{L}$.
This term involves the expansion parameter $\Theta$ along the null geodesics,
which will be introduced in eq. (\ref{expansion-parameter}).
\end{itemize}
Since the geometries under consideration are static and do not present causally disconnected boundaries,
it is not restrictive to consider the case where the time on both boundaries is vanishing.
In order to set up the actual computation, we need to determine the WDW patch
 and the EW, which both require to analyze null geodesics in the spacetime of interest.
Before doing that, we comment on the regularization prescriptions that will be used throughout the paper.

\subsection{Regularization prescriptions for UV divergences}
\label{sect-regularization_prescriptions}

We are interested in the computation of the UV divergences of the subregion action for theories with defects.
These kinds of geometries can be described by performing an $\mathrm{AdS}_d$ slicing
 of asymptotically $\mathrm{AdS}_{d+1}$ space to obtain a metric in the form \cite{Estes:2014hka}
\beq
ds^2 = L^2 \le  A^2 (y) ds^2_{\mathrm{AdS}_d} + \rho^2(y) dy^2 \ri \, ,
\label{eq:metric_Trivella_form}
\eeq
where $y$ is a non-compact coordinate such that when $y \rightarrow \pm \infty$ 
\beq
A(y) \rightarrow \frac{L_{\pm}}{2} e^{\pm y \pm c_{\pm}} \, , \qquad
\rho(y) \rightarrow 1 \, ,
\label{eq:asymptotic_Trivella_functions}
\eeq
where $L_{\pm}$ and $c_{\pm}$ are constants.
We parametrize the $\mathrm{AdS}_d$ slices using Poincaré coordinates 
\beq
ds^2_{\mathrm{AdS}_d} = \frac{1}{z^2} \le dz^2 - dt^2 + d \vec{x}^2_{d-2} \ri \, ,
\label{eq:metric_AdS_slicing}
\eeq
where $(t,z)$ are the time and radial coordinates on each slice and $\vec{x}$ collects all the other orthogonal directions.

Three different regularisation prescriptions have been used the literature \cite{Estes:2014hka,Bak:2016rpn,Gutperle:2016gfe}:
\begin{itemize}
\item The Fefferman-Graham (FG) regularization 
 relies on performing
 a FG expansion of the metric to select a radial direction $\xi$ for the asymptotic
  AdS$_{d+1}$ region in Poincaré coordinates, and introducing a UV cutoff by cutting the spacetime with a surface located at $\xi=\delta.$
The metric in FG form reads
\beq
ds^2 = \frac{L^2}{\xi^2} \left[ d \xi^2 + g_1 (\xi/\eta) \,  \le -dt^2 + d\vec{x}^2 \ri + g_2 (\xi/\eta) \, d\eta^2  \right] \, ,
\label{eq:form_metric_FGexpansion}
\eeq
where $\xi$ is a radial coordinate for the asymptotic AdS region in Poincaré coordinates,
 $\eta$ is the boundary direction orthogonal to the defect, and $g_1, g_2$ are two appropriate functions,
 such that  the original metric \eqref{eq:metric_Trivella_form} with slicing \eqref{eq:metric_AdS_slicing}
 is equivalent to (\ref{eq:form_metric_FGexpansion}) with a suitable change of coordinates $(z,y) \rightarrow (\xi, \eta).$ 
In the region $\xi \gg \eta$, the FG expansion breaks down because the 
coordinates $\xi$ and $\eta$ are not well defined \cite{Papadimitriou:2004rz}.
This problem can be solved  
by introducing a continuous curve which interpolates between the right and left patches of the defect  \cite{Estes:2014hka}.
\item In the single cutoff regularization \cite{Bak:2016rpn} the arbitrary interpolation curve is replaced
by a cutoff on the minimal value of the $z$ coordinate such that
\beq
 \delta=\frac{z}{A(y)}  \, , \qquad
z_{\rm min} = \delta \, \underset{y \in \mathbb{R}}{\mathrm{min}} \, [A(y)] \, .
\label{eq:single_cutoff_prescription}
\eeq
The physical quantities are then expanded in series around $\delta=0.$ 
\item The double cutoff regularization \cite{Gutperle:2016gfe}  introduces two different cutoffs 
for each of the directions $(y,z)$.
The first cutoff is directly imposed on the $\mathrm{AdS}_d$ slicing at $z= \delta.$
The second cutoff regularizes the divergences of $A(y)$ at infinity. 
This can be achieved by restricting the $y$ domain up to a maximum value  $y^*$, defined by
\beq
A(y^*) = \frac{1}{\varepsilon} \, .
\label{eq:double_cutoff_prescription} 
\eeq 
While the $\delta$ cutoff has physical relevance since it regularizes the intrinsic contributions from the defect, 
the $\varepsilon$ cutoff is only a mathematical artifact introduced at intermediate steps. 
Observables which are intrinsic to the defect must be $\varepsilon$-independent after the subtraction of the vacuum solution.
\end{itemize}
The advantage of both the FG and the single cutoff regularizations is the introduction of only one regulator; the drawback is that integrals along different coordinates are nested.
From a technical point of view, it is simpler to consider the double cutoff regularization.
For volume complexity, we checked \cite{Auzzi:2021nrj, Baiguera:2021cba} that all the three methods only differ by  finite parts, while we expect universal contributions to appear in logarithmically divergent terms.
We then choose to evaluate the action complexity with the double cutoff method, since the universal behaviour is not influenced by the regularization scheme.

The previous discussion applies in particular to the case of the Janus AdS geometry and will be used in section \ref{sect-computation_action_Janus}.
The case of the AdS/BCFT model is simpler, because the spacetime is empty AdS space with the addition of an end-of-the-world brane.
While it is still possible to employ the parametrization \eqref{eq:metric_Trivella_form}, it is simpler to work using three-dimensional Poincaré coordinates including a UV regulator cutting the spacetime with the surface $z=\delta,$ without any need for a second parameter $\varepsilon.$
This choice will be used for the computations in section \ref{sect-computation_action_BCFT}.

\begin{figure}[ht]
\centering
\includegraphics[scale=0.17]{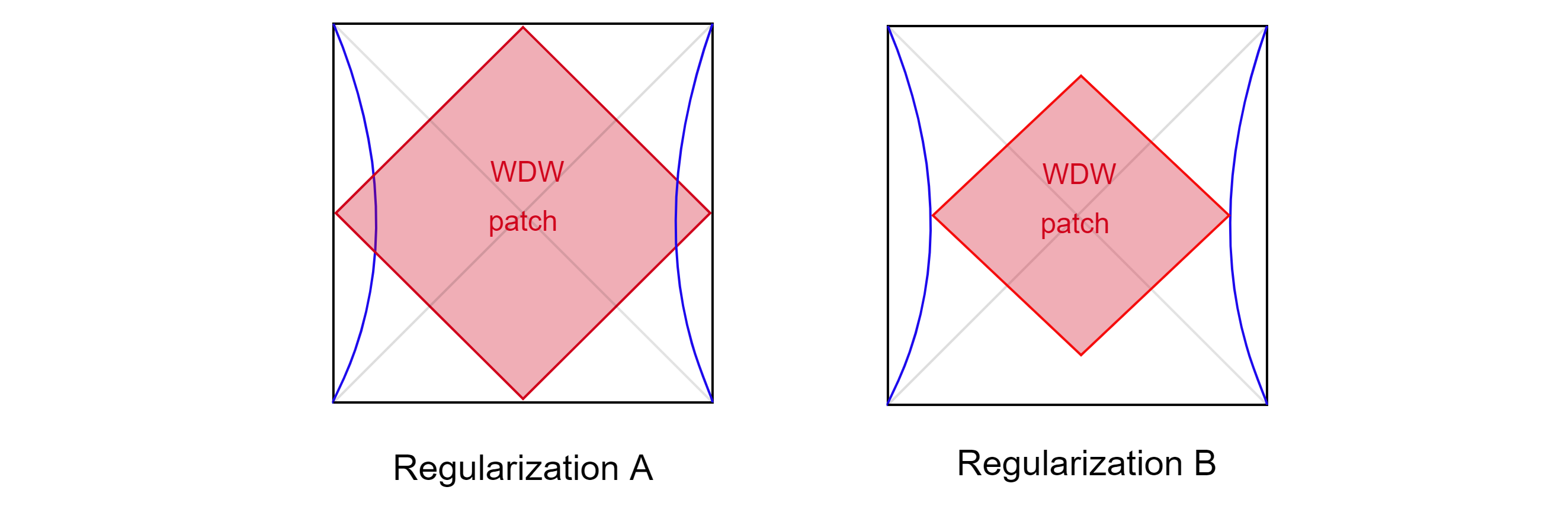}
\caption{The two regularizations of the WDW patch  introduced in \cite{Carmi:2016wjl}.
In the picture it is represented the case of a black hole in asymptotically AdS spacetime.}
\label{fig-2regs}
\end{figure}

The three regularization procedures discussed above do not represent the only ambiguities involving the computation of the action.
It is also possible to define the WDW patch surfaces in two different ways \cite{Carmi:2016wjl}, depicted in Fig.~\ref{fig-2regs}:
\begin{itemize}
\item Regularization A amounts to build the WDW patch starting from the true boundary located at $z=0,$ and cut the spacetime with a surface located at $z=\delta.$ 
\item Regularization B corresponds to the null geodesics delimiting the WDW patch to directly start from the cutoff surface surface located at $z=\delta.$ 
\end{itemize}
The same ambiguity arises for the  Ryu-Takayanagi (RT) surface.

In this paper we will  work with regularization A.
The divergences of regulatization B are related to the ones of regularization A by another counterterm that can be introduced on the cutoff at the boundary \cite{Akhavan:2019zax, Omidi:2020oit}, in the spirit
of holographic renormalization \cite{Balasubramanian:1999re,deHaro:2000vlm,Bianchi:2001kw,Skenderis:2002wp}.
We will check that in our case such a counterterm is finite (see Appendix \ref{app-timelike_counterterm}).

\section{Subregion complexity in the Janus AdS$_3$ geometry}
\label{sect-complexity-janus}

In this section we perform the computation of the subregion action complexity in the Janus AdS$_3$ spacetime. 
In section  \ref{sect-null_geodesics_conf_rescaling}
we review some background material on null surfaces and on the properties
of null geodesics under conformal transformations.
In section \ref{sect-null_boundaries_Janus_AdS} we find the integration domain 
for the action. In section \ref{sect-computation_action_Janus}
we perform the actual calculation.

\subsection{Null surfaces, geodesic congruencies and conformal rescalings}
\label{sect-null_geodesics_conf_rescaling}

The boundary of the integration domain of the action will involve many null hypersurfaces.
Let us review a few useful properties, following \cite{Poisson:2009pwt}.
We consider a null  hypersurface $\Sigma$ selected by an appropriate restriction on the coordinates  $x^{\mu}$ of spacetime  
\beq
\Phi(x^{\mu})=0  \, ,
\label{surfe}
\eeq
where $\Phi$ is a scalar function increasing towards the future.
The normal one-form $k_{\mu}$ to $\Sigma$, defined by
\beq
k_{\mu}=- \alpha \p_{\mu} \Phi \, ,
\eeq
is by construction null $k^{\mu} k_{\mu}=0$ and also tangent to $\Sigma$.
Here $\alpha>0$ is a constant prefactor, such that the corresponding vector $ k^{\mu}$ is future-oriented.

Moreover, the vector field $k^{\mu}$ satisfies the geodesic equation 
\beq
k^{\nu} D_{\nu} k^{\mu} = \kappa \, k^{\mu} \, ,
\label{eq-geo-geo}
\eeq
where $\kappa$ is a spacetime scalar which vanishes for  affine parameterisations.

We can in general parameterize
a null hypersurface as a congruence of null geodesics.
Besides using an implicit expression of kind (\ref{surfe}) 
an alternative way to parametrize a null hypersurface is through the expression $X^{\mu}=X^{\mu} (\lambda, \theta^A)$
 with the requirements that 
 the parameter $\lambda$ moves along a single generator in the congruence of null geodesics, 
 and  the parameters $\theta^A$ are constant on each null generator spanning the hypersurface.
We define then the tangent vectors along the hypersurface to be
\beq
k^{\mu} \equiv \frac{d X^{\mu}}{d \lambda} \, , \qquad
e^{\mu}_A \equiv  \frac{d X^{\mu}}{d \theta^A} \, , 
\label{eq:data_congruence_null_geodesics}
\eeq
where $k^{\mu}$ is the null tangent vector and $e^{\mu}_A$ is a spacelike vector, defined in such a way
that it is orthogonal to $k^\mu$, i.e.
\beq
k^{\mu} e^A_{\mu} = 0 \, .
\label{eq:orthogonality_constraint_null_vector_vielbein}
\eeq
The normal $k^\mu$ satisfies eq. (\ref{eq-geo-geo}) where $\kappa$ is a function of $\lambda$.
The vectors $e^\mu_A$ define the induced metric
\beq
\gamma_{AB} = g_{\mu\nu} e^{\mu}_A e^{\nu}_B \, .
\eeq
The expansion parameter along the congruence of null geodesics is given by
\beq
\Theta = \frac{1}{\sqrt{\gamma}} \frac{d \sqrt{\gamma}}{d\lambda} \, ,
\label{expansion-parameter}
\eeq
where $\gamma$ is the determinant of $\gamma_{AB}$.

In order to determine the null hypersurface in Janus,
it will be convenient to perform a conformal rescaling. 
Let us review a few basic properties \cite{Blau,Wald:1984rg}
Consider two metrics $g$ and $\tilde{g}$ related by a conformal transformation
\beq
\tilde{g}_{\mu\nu}(x^{\prime})=\Omega(x)^2 \, g_{\mu\nu}(x) \, .
\label{eq:conformal_factor}
\eeq
Under this map the causal structure of the spacetime is preserved.
While in general the two metrics have different spacelike and timelike geodesics,
the null geodesics are the same. However, the affine parameterization condition of these geodesics
in general is not preserved under (\ref{eq:conformal_factor}).

As a matter of fact, starting from an affine parameterization for the geodesics of the metric $g$, i.e.
\beq
k^{\nu} D_{\nu} k^{\mu} = 0 \, ,
\label{eq:geodesic_eq_rescaled_metric}
\eeq
under the conformal transformation (\ref{eq:conformal_factor}) the equation (\ref{eq:geodesic_eq_rescaled_metric})
becomes
\beq
\label{modified_geodesic_equations}
k^{\nu} \tilde{D}_{\nu} k^{\mu}=\kappa \, k^{\mu} \, , \qquad
\kappa=\frac{2}{\Omega}\frac{d\Omega}{d\lambda} \, .
\eeq
We can think of $\kappa(\lambda)$ as the measure of the failure of $\lambda$ to be
an affine parameter.

%

\subsection{Null boundaries in the Janus AdS$_3$ geometry}
\label{sect-null_boundaries_Janus_AdS}

Using a conformal transformation with
\beq
\label{conformal_factor}
\Omega^2= L^2 \frac{f(\mu)}{z^2}\;.
\eeq
we can write the Janus metric 
 \eqref{eq:metric_Janus_mucoord} as 
 \beq
 d\tilde{s}^2=\Omega^2 \, ds^2 \, , \qquad ds^2=-dt^2+dz^2 + z^2d\mu^2 
 \label{Janus_metric_conformal_factor}
 \eeq
 where $ds^2$ is the flat spacetime metric in polar coordinates.
 We proceed to study the null congruence of geodesics delimiting the WDW patch and the EW.
 
%
%

\vskip 4mm \noindent
{\bf {WDW patch.}}
By going to cartesian coordinates
\beq
X= z \sin \mu \, , \qquad
Y= z \cos \mu \, ,
\eeq
we bring the metric to the form
$ ds^2 = -dt^2 + dX^2 + dY^2$. 
In these coordinates, the two-dimensional plane specified by
\beq
g(t,X,Y) = a \, X + b \, Y + c \, t = 0 \, , \qquad c = \pm \sqrt{a^2 + b^2} \, ,
\label{gg-surface}
\eeq
is a null surface.
From now on we will specialize to the case of $c<0,$ in view of the parametrization of the part at positive time of the WDW patch.

From the general result that null geodesics are invariant under conformal transformations
of the metric, it follows that eq. (\ref{gg-surface}) specifies a null surface also
 in the Janus $\mathrm{AdS}_3$ background. In polar coordinates, it reads
\beq
g(t,\mu,z) = a \, z \sin \mu + b \, z \cos \mu + c \, t = 0 \, .
\eeq
Let us first consider  the WDW patch anchored at the right boundary $\mu=\mu_0.$
In order to impose that this null surface is part of the 
boundary of the WDW patch in the regularization A, we enforce that at $t=0,$ 
the angular coordinate is $\mu= \mu_0.$ 
This condition determines
\beq
 \tan \mu_0 = - \frac{b}{a} \, .
\eeq
Therefore we conclude that the null boundary of the WDW patch is described by the equations
\beq
t_{\rm WDW} (\mu,z) =   z \sin \le \mu_0 - \mu \ri \, ,
\label{eq:expression_tWDW_positive_time_simplified}
\eeq
where we used as a working assumption that $\mu_0 \in [\pi/2, \pi].$ 
Using eq.~(\ref{eq:data_elliptic_functions}), this implies
 \beq
 \gamma  \leq \gamma_0 \approx 0.704 \, .
 \label{gamma-zero-assumption}
 \eeq 
For simplicity, we will restrict to the case $\gamma \in [0,\gamma_0]$.
In the case $\gamma \in  \left( \gamma_0, \frac{1}{\sqrt{2}} \right),$ 
the geometry of the WDW patch changes.


We can then describe the boundary of the WDW patch as
\beq
X^{\mu}_{\rm WDW} = \le  t_{\rm WDW} (\mu,z) , \mu , z \ri \, .
\eeq
Since the spacetime is three-dimensional, there is only one coordinate $\theta$ entering eq.~\eqref{eq:data_congruence_null_geodesics}, i.e.
\beq
k^{\mu}_R = \frac{dX^{\mu}_{\rm WDW}}{d \lambda} \, ,
\qquad
e^{\mu}_R = \frac{dX^{\mu}_{\rm WDW}}{d \theta} \, .
\label{eee-kappone}
\eeq
We should also impose that the
 orthogonality condition \eqref{eq:orthogonality_constraint_null_vector_vielbein} holds.
 The affine parameterization is not convenient for the calculation\footnote{The reason is technical: the affine parameter is determined by the integral curves generated by the vector field $k^{\mu},$ but the differential equation contains the conformal factor, which depends on $f(\mu)$ defined in eq.~\eqref{eq:solutions_f_dilaton_mu_coordinates}. The choice we adopt in the main text avoids the appearance of such conformal factor in the differential equation.}.
 It turns out that a convenient choice is
\beq
\lambda = - \frac{z}{\alpha} \sin \le \mu_0 - \mu \ri \, , \qquad
\theta = - \log \le z \cos \le \mu_0 - \mu \ri  \ri \, ,
\label{eq:parametrization_WDW_patch_Janus_AdS}
\eeq
where $\alpha>0$ parametrizes the ambiguity in the normalization of a null vector.
With this choice, eq. (\ref{eee-kappone}) takes the form
\beq
k^{\mu}_R = - \alpha \le 1, -\frac{\cos \le \mu_0 - \mu \ri}{z}, \sin \le \mu_0 - \mu \ri \ri \, , 
\eeq
\beq
e^{\mu}_R = - \le 0, \frac{1}{2} \sin \le 2 (\mu_0 - \mu) \ri, z \cos^2 \le \mu_0 - \mu \ri \ri  \, ,
\eeq
The induced metric $\gamma$ (which is a number because it is a $1$ by $1$ matrix), the expansion parameter $\Theta$
and the scalar $\kappa$ are given by:
\bea
\gamma_{\rm WDW}  &=& L^2 f(\mu) \cos^2 \le \mu_0 - \mu \ri \, ,
\nl
\Theta_{\rm WDW} &=& \alpha \, \left[ \frac{\sin \le \mu_0 - \mu \ri}{z} + \frac{\cos \le \mu_0 - \mu \ri}{2 z} \frac{f'(\mu)}{f(\mu)} 
\right] \, ,
\nl 
\kappa_{\rm WDW} (\lambda) &=&   2 \Theta_{\rm WDW} \, .
\label{gamma-theta-kappa-left}
\eea

The boundary of the WDW patch anchored at the left boundary $L$ (placed at $\mu=-\mu_0$)
can be treated in an analog way\footnote{Notice that the left side of the WDW patch with positive times is obtained from the right side by sending $\mu \rightarrow - \mu .$ We apply this change in the choice of the parametrization as well.}, obtaining
\beq
X^{\mu}_{\rm WDW,L}  = \le z \sin \le \mu + \mu_0 \ri, \mu, z \ri \, .
\eeq 
The corresponding parameterization is
\beq
\lambda = - \frac{1}{\alpha} \, z \sin \le \mu + \mu_0 \ri \, , \qquad
\theta = - \log \le z \cos \le \mu + \mu_0 \ri \ri \, , 
\eeq
which gives the following vectors
\bea
k^{\mu}_L &=& -\alpha \le 1, \frac{\cos \le \mu + \mu_0 \ri}{z}, \sin \le \mu + \mu_0 \ri \ri \, , 
\nl
e^{\mu}_L &=&  \le 0, \frac{1}{2} \sin \le 2 (\mu + \mu_0) \ri, -z \cos^2 \le \mu + \mu_0 \ri \ri  \, ,
\eea
and the following geometric data
\bea
\gamma_{\rm WDW,L}  &=& L^2 f(\mu) \cos^2 \le \mu + \mu_0 \ri \, ,
\nl
\Theta_{\rm WDW,L} &=& \alpha \, 
\left[ \frac{\sin \le \mu + \mu_0 \ri}{z} - \frac{\cos \le \mu + \mu_0 \ri}{2 z} \frac{f'(\mu)}{f(\mu)} 
\right] \, ,
\nl
\kappa_{\rm WDW,L} (\lambda) &=&   2 \Theta_{\rm WDW,L}  \, .
\eea
For future convenience, we write the normal to the left and right side of the WDW patch
as one-forms
\bea
\label{eq:normal_one_form_WDW_patch_Janus}
\mathbf{k}_R = \alpha \, \frac{L^2 f(\mu)}{z^2} \le dt + z \cos \le \mu_0 - \mu \ri d\mu - \sin \le \mu_0 - \mu \ri dz  \ri \, ,
\nl
\mathbf{k}_L =  \alpha \, \frac{L^2 f(\mu)}{z^2} \le dt - z \cos \le \mu + \mu_0 \ri d\mu - \sin \le \mu + \mu_0 \ri dz  \ri   \, .
\eea


\vskip 4mm  \noindent
{\bf{Entanglement wedge.}}
Since the Janus metric is conformally equivalent to $2+1$ dimensional
flat spacetime, the two spaces share the same null geodesics,
in particular the lightcones at constant $\mu$
\beq
\label{null_geodesics-const-y}
z=\pm t \pm c \, .
\eeq
Here the constant $c$ will be determined by suitable boundary conditions.
We will show that the geodesics (\ref{null_geodesics-const-y})
 define the boundary of the EW. 
 
 The Ryu-Takayanagi (RT) surface anchored at the boundary \cite{Azeyanagi:2007qj} is described
 by the equation
 \beq
 z_{\rm RT} = l/2 
 \eeq 
for an interval of length $l$ located symmetrically along $\mu$ on the surface at constant time $t=0.$
By imposing that the curves at constant $\mu$ given in eq.~\eqref{null_geodesics-const-y} pass through
 the RT surface, we determine that  the null boundary of the EW is
 \beq
t_{\rm EW} = \frac{l}{2} - z \, ,
\label{eq:curve_EW}
\eeq
where we are restricting the solution to the part with positive time coordinate.
This expression holds both in empty AdS space and in the Janus background. 

It is convenient to work in the affine parametrization.
The Langrangian which describes affinely parametrized geodesics is of the form
\beq
\label{Lagrangian_for_null_geodesics}
\mathcal{L}=\Omega^2 \le -\dot{t}^2+\dot{z}^2 +z^2\dot{\mu}^2 \ri \, ,
\eeq
where  dot denotes derivative with respect to the affine parameter.
The equations of motion give the following tangent vector for the null geodesics with constant $\mu$ 
\beq
w^\mu=\b \, (\dot{t}, \dot{z},0) =\b  \frac{1}{\Omega^2} (-1,1,0) \, ,
\eeq
where $\b$ is an arbitrary constant.
Lowering the indices, we get the one-form
\beq
\mathbf{w} = \beta \le  dt + dz \ri  \, .
\label{eq:null_one_form_EW}
\eeq
Note that the dependence on the conformal factor disappears
on the form $\mathbf{w}$.
Such one-form is orthogonal at the boundary to the curve parametrizing the RT surface, \emph{i.e.}
\beq
w_{\mu} \frac{d X^{\mu}_{\rm RT}}{d \lambda}\Big|_{\rm bdy} = 0 \, ,
\label{eq:parametrization_null_oneforms_EW}
\eeq
This shows that the congruence of null geodesics \eqref{eq:curve_EW} describes indeed the null boundary of the EW.
As anticipated, the parametrization is affine and therefore $\kappa_{\rm EW} = 0$.
From eq.~(\ref{expansion-parameter}), we find that the expansion parameter vanishes $ \Theta_{\rm EW} = 0$,
as expected on general grounds since the EW is delimited by an extremal surface \cite{Headrick:2014cta}.

\begin{figure}[ht]
\centering
\includegraphics[scale=0.5]{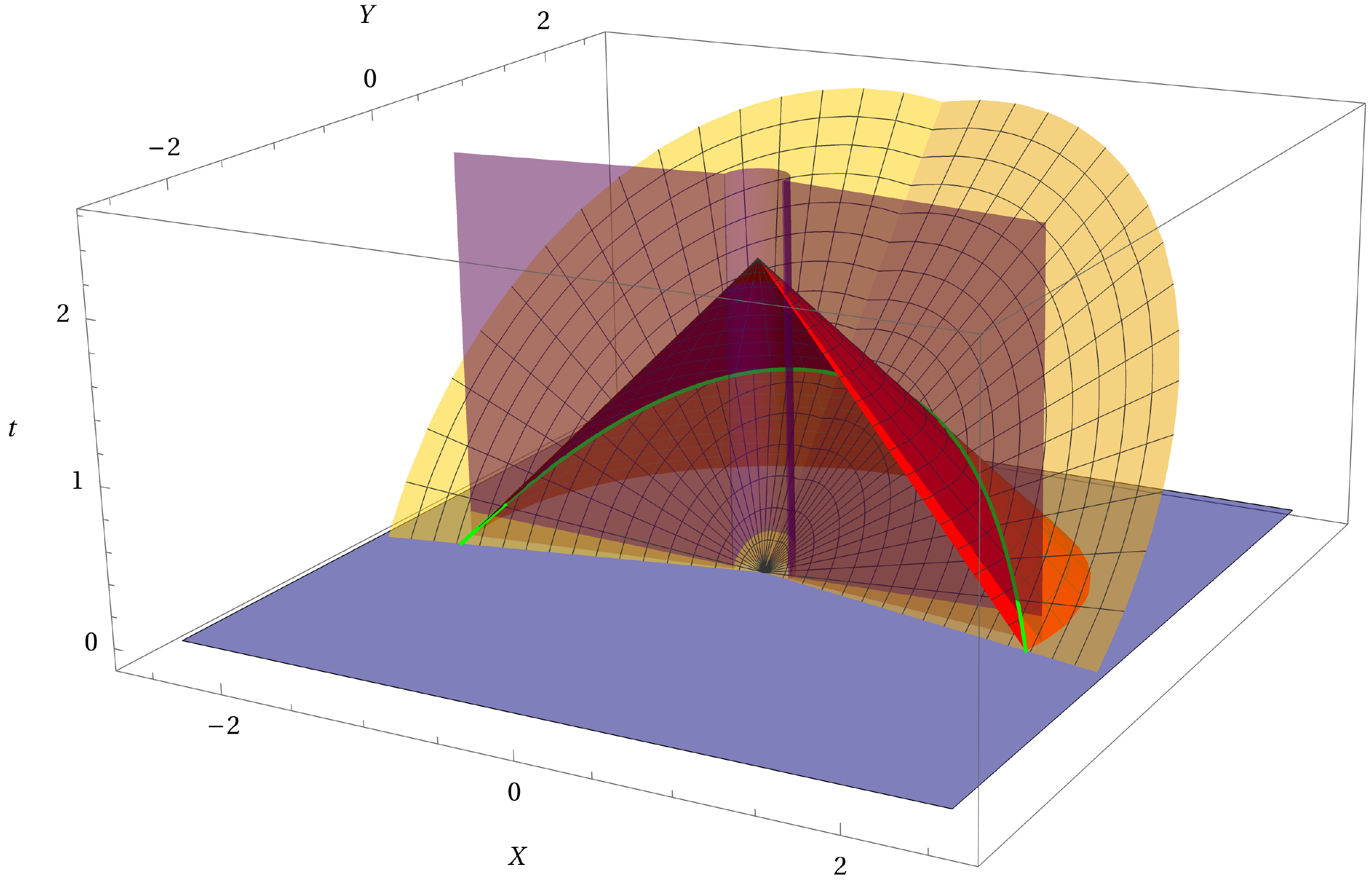}
\caption{A pictorial representation of the null boundaries of the WDW patch (light orange surfaces), 
of the entanglement wedge (red), and the intersection curve $z_{\rm int} (\mu)$ between them (green). 
The blue plane restricts the integration region to $t\geq 0$ only.
The purple transparent region represents the cutoff surfaces 
located at $z=\delta$ and $\mu= \pm \mu^* (\varepsilon).$
The diagram corresponds to the choice $\gamma= 0.5.$}
\label{fig-geometrical_configuration_action}
\end{figure}

\vskip 4mm \noindent
{\bf{Intersection curve.}} In view of the computation of the gravitational action, we need to determine the intersection curve between the WDW patch and the EW.
It is sufficient for symmetry reasons to focus on the region with positive $(t,\mu).$
By equating the hypersurfaces defined in
eq.~(\ref{eq:expression_tWDW_positive_time_simplified})
and eq.~(\ref{eq:curve_EW}),
we obtain the following curve
\beq
z_{\rm int} (\mu) = \frac{l}{2} \frac{1}{\sin \le \mu_0 - \mu \ri +1} \, .
\label{eq:intersection_curve_Janus_AdS}
\eeq
A picture of the WDW patch, the EW and their 
intersecton curve in $(t,X,Y)$ coordinates is shown in Fig.~\ref{fig-geometrical_configuration_action}.

\subsection{Computation of the action}
\label{sect-computation_action_Janus}

We are now ready to compute the gravitational action \eqref{eq:total_gravitational_action}
 in the Janus AdS$_3$ background using the double cutoff prescription.
The equivalent of eq.~\eqref{eq:double_cutoff_prescription} for this case is 
\beq
\sqrt{f(y^*)} = \frac{1}{\varepsilon} \, .
\label{eq:def_cutoff_Janus_AdS}
\eeq 
This equation determines a value of $y^*(\varepsilon)$ which delimits the corresponding integration
\beq
y^* (\varepsilon)= \frac{1}{2} \, \mathrm{arccosh}  \le \frac{\frac{2}{\varepsilon^2}-1}{\sqrt{1-2\gamma^2}} \ri \, .
\label{eq:definition_ystar}
\eeq
We can express this result 
 in terms of the $\mu$ coordinates by means of the change of variables
\beq
\tanh y = \mathrm{sn} \le \alpha_+ \mu | m \ri \, ,
\label{tanh-sn-relazione}
\eeq 
which infinitesimally corresponds to  eq.~\eqref{eq:Janus_metric}
and where we are using the definitions \eqref{eq:data_elliptic_functions}.
A derivation of eq. (\ref{tanh-sn-relazione}) is provided in in eq. (\ref{derivazione-eq-3.40}).
The corresponding value of the cutoff in the $\mu$ variable,
such that $\sqrt{f(\mu^*(\varepsilon))} = \varepsilon^{-1}$, can be obtained by combining 
 eqs.~\eqref{eq:definition_ystar} and ~\eqref{tanh-sn-relazione}.

We compute the subregion action term by term.
Our general strategy will be the following: we will evaluate explicitly the integrations over $(t, z),$ while 
we will collect all the integrands in the $\mu$ 
variables, and extract their divergences at the very end of the calculation.

\subsubsection{Bulk term}

The bulk term reads
\beq
I_{\mathcal{B}} = \frac{1}{16 \pi G} \int_{\rm WDW \cap EW} d^3 x \, \sqrt{-g}	 \,  \mathcal{L}\, ,  \qquad
\mathcal{L}=  R + \frac{2}{L^2}  - g^{\mu\nu} \p_{\mu} \phi \p_{\nu} \phi  \, ,
\label{eq:general_bulk_action}
\eeq
where $R$ is the Ricci scalar of the metric \eqref{eq:Janus_metric} with profile function $f$ and dilaton solution given in eq.~\eqref{eq:function_f_and_dilaton}.
It is important to remark that the presence of the defect is responsible for the backreaction of the original vacuum $\mathrm{AdS}_3$ spacetime, which leads to a different value of the Ricci scalar than empty AdS.
However, the addition of a kinetic term for the dilaton gives a simple  on-shell action which reads
\beq
\mathcal{L} 
 = - \frac{4}{L^2} \, .
\eeq
The intersection curve between the null boundaries of the WDW patch and the EW naturally splits the integration region in two parts:
\beq
I_{\mathcal{B}} = I_{\mathcal{B}}^1 + I_{\mathcal{B}}^2  \, ,
\eeq
where
\bea
I_{\mathcal{B}}^1 &=& - \frac{L}{\pi G} \int_0^{\mu^* (\varepsilon)} d\mu  \int_{\delta}^{z_{\rm int} (\mu)} dz
  \int_0^{t_{\rm WDW} (\mu,z)} dt \,  \frac{f(\mu)^{3/2}}{z^2} \, ,
\nl
I_{\mathcal{B}}^2 &=& - \frac{L}{\pi G} \int_0^{\mu^* (\varepsilon)} d\mu \int_{z_{\rm int} (\mu)}^{z_{\rm RT}} dz 
\int_0^{t_{\rm EW} (z)} dt \,  \frac{f(\mu)^{3/2}}{z^2} \, .
\eea
We introduced a symmetry factor of 4 coming from the integrations along $(t,\mu).$
A direct evaluation of the integrals over $(t,z)$ brings to the result
\bea
I_{\mathcal{B}} = \frac{L}{\pi G} \int_0^{\mu^*(\varepsilon)} d\mu \, f(\mu)^{3/2} \,
 \biggl\{ \le \sin \le \mu_0 - \mu \ri +1 \ri \log \left[  \sin \le \mu_0 - \mu \ri +1 \right]  
\biggr.
\nl
 \biggl.
+ \sin \le \mu_0 - \mu \ri \left[ \log \le \frac{2 \delta}{l} \ri -1  \right]  \biggr\} \, .
\label{eq:bulk_term_JAdS_before_exp}
\eea

\subsubsection{GHY term}

The regularization prescription A in Fig.~\ref{fig-2regs} requires to evaluate
the GHY term 
\beq
I_{\rm GHY} = \frac{\varepsilon_{t,s}}{8 \pi G} \int_{\mathcal{B}_{t,s}} d^2 x \, \sqrt{|h|} \, K \, ,
\label{azione-gibbons-hawking-york}
\eeq
at the cutoff surfaces $z=\delta$ and $\mu= \pm \mu^* (\varepsilon).$
In eq.~(\ref{azione-gibbons-hawking-york})
 $K$ is the trace of the extrinsic curvature, $h$ the determinant of the induced metric,
  and $\varepsilon_{t,s}=\pm 1$ if the surface of interest ${\mathcal{B}_{t,s}} $ is timelike or spacelike, respectively.
 
\vskip 4mm \noindent
{\bf Cutoff surface located at $z=\delta$.} 
The unit normal vector $n^{\mu}$ to the cutoff surface located at $z=\delta$ is given by
\beq
n_1^{\mu} = - \frac{z}{L \sqrt{f(\mu)}} \le 0, 0, 1 \ri \, ,
\label{eq:normal_n1_JAdS}
\eeq
the minus sign being chosen in such a way that it is outward-directed  from the region of interest for the computation of the action.
The determinant of the induced metric and the trace of the extrinsic curvature are
\beq
\sqrt{-h} = L^2 \, \frac{f(\mu)}{z} \, ,
\qquad
K = D_{\mu} n_1^{\mu} = \frac{1}{L \sqrt{f(\mu)}} \, .
\eeq
In principle, one would expect to split the integration region according
 to the intersection curve $z_{\rm int} (\mu)$
 in eq. (\ref{eq:intersection_curve_Janus_AdS})
  between the null boundaries of the WDW patch and of the EW.
However, a choice of a small enough value of $\delta$ always allows 
to keep the entire cutoff surface inside the region where the WDW patch sits below the EW, since
\beq
z_{\rm int} (\mu) = \frac{l}{2} \frac{1}{\sin \le \mu_0 - \mu \ri +1} \geq \frac{l}{4} \, .
\eeq
For this reason, the GHY term reads
\beq
I^{\delta}_{\rm GHY}  =
 \frac{L}{2 \pi G} \int_0^{\mu^* (\varepsilon)} d\mu \int_0^{t_{\rm WDW} (\delta, \mu)} dt \, \frac{\sqrt{f(\mu)}}{\delta} \, ,
\eeq
where we put a symmetry factor of 4 
(there is a factor of 2 from each of the two integrations).
Performing the first integration, we find
\beq
I^{\delta}_{\rm GHY} = \frac{L}{2 \pi G} \int_0^{\mu^* (\varepsilon)} d\mu \, \sqrt{f(\mu)} \, \sin \le \mu_0 - \mu \ri \, .
\label{eq:GHY_term1_JAdS_before_exp}
\eeq

\vskip 4mm \noindent
{\bf Cutoff surface located at $\mu=\mu^*(\varepsilon)$.} 
The outward-directed normal to the cutoff surface located at $\mu=\mu^* (\varepsilon)$ is given by
\beq
n_2^{\mu} = \frac{1}{L \sqrt{f(\mu)}} (0,1,0) \, ,
\label{eq:normal_n2_JAdS}
\eeq
and the determinant of the induced metric and the trace of the extrinsic curvature  are
\beq
\sqrt{-h} = L^2 \frac{f(\mu)}{z^2} \, ,
\qquad
K = D_{\mu} n^{\mu}_2 = \frac{1}{L} \frac{f'(\mu)}{f(\mu)^{3/2}} \, .
\eeq
In this case, the surface $\mu= \mu^* (\varepsilon)$ cuts both the WDW patch and the EW; for this reason, the GHY term
 decomposes into two parts determined by the intersection curve in eq.~\eqref{eq:intersection_curve_Janus_AdS}.
These codimension-one boundary terms are given by
\beq
I^{\varepsilon}_{\rm GHY} = I^{\varepsilon}_{\rm GHY,1} + I^{\varepsilon}_{\rm GHY,2} \, ,
\eeq
where
\bea
 I^{\varepsilon}_{\rm GHY,1}  &=&  \frac{L}{2 \pi G} \int_{\delta}^{z_{\rm int} (\mu^* (\varepsilon))}  dz  \int_0^{t_{\rm WDW} (z,\mu^*(\varepsilon))} dt \, 
 \frac{f'(\mu)}{z^2 \, \sqrt{f(\mu)}} \Big|_{\mu=\mu^*(\varepsilon)} 
\nl
 I^{\varepsilon}_{\rm GHY,2}  &=&  \frac{L}{2 \pi G} \int_{z_{\rm int} (\mu^* (\varepsilon))}^{z_{\rm RT}}  dz  \int_0^{t_{\rm EW} (z)} dt \, 
 \frac{f'(\mu)}{z^2 \, \sqrt{f(\mu)}} \Big|_{\mu=\mu^*(\varepsilon)}  
\eea
In these integratons, we already put the symmetry factor of 4 (one factor of 2 comes from the integration along $t,$ while the other factor of 2 arises because we also account for the cutoff surface located at $\mu=-\mu^*(\varepsilon)$ by symmetry reasons).
Both these integrations can be performed explicitly, and further simplifications occur by using
 properties of the elliptic functions  to find
\beq
f \le \mu^* (\varepsilon) \ri = \frac{1}{\varepsilon^2} \, , \qquad
\frac{df}{d\mu}\Big|_{\mu=\mu^*(\varepsilon)} = \frac{2}{\varepsilon^3} \sqrt{\le 1- \alpha_+^2 \varepsilon^2 \ri \le 1- m \alpha_+^2 \varepsilon^2  \ri} \, .
\eeq
Employing these identities and solving the integrals, we find
\beq
\begin{aligned}
I^{\varepsilon}_{\rm GHY} = & \frac{L}{\pi G \varepsilon^2}  \sqrt{\le 1- \alpha_+^2 \varepsilon^2 \ri \le 1- m \alpha_+^2 \varepsilon^2  \ri}  \,  
\left\lbrace \sin \le \mu_0 - \mu^*(\varepsilon) \ri \left[1- \log \le \frac{2 \delta}{l} \ri   \right]  \right. \\
& \biggl.  - \left[ \sin \le \mu_0 - \mu^*(\varepsilon) \ri +1  \right]  \, \log \left[ \sin \le \mu_0 - \mu^*(\varepsilon) \ri +1  \right] \biggr\} \, .
 \end{aligned}
 \label{eq:GHY_term2_JAdS_before_exp}
\eeq

\subsubsection{Null boundary terms}

The contribution due to null boundaries is of the following form
\beq
I_{\mathcal{N}} = \frac{\varepsilon_n}{8 \pi G} 
\int_{\mathcal{B}_n} d\lambda d \theta \, \, \sqrt{\gamma} \, \kappa (\lambda) \, ,
\label{termine-azione-bordi-null}
\eeq
where $\varepsilon_n = \pm 1$ depends on the orientation of the null normal to the surface, 
$\lambda$ is a parameter along the congruence of geodesics, 
$\gamma$ is the induced metric along the $\theta$ direction and $\kappa(\lambda)$ is defined in the geodesic equation (\ref{eq-geo-geo}).

\vskip 4mm  \noindent
{\bf WDW patch.} Using the parametrization in eq.~\eqref{eq:parametrization_WDW_patch_Janus_AdS}, 
the WDW patch is not affinely parametrized and therefore we need to evaluate the following integral:
\beq
I^{\rm WDW}_{\mathcal{N}} = \frac{\alpha L}{4 \pi G} \iint d \lambda \, d \theta \, 
 \, \left[ \frac{\sin \le \mu_0 - \mu \ri}{z} + \frac{\cos \le \mu_0 - \mu \ri}{2 z} \frac{f'(\mu)}{f(\mu)} \right]  \, \sqrt{f(\mu)} \left| \cos \le \mu_0 - \mu \ri \right| \, .
\eeq
We take $\varepsilon_n = 1$ because the spacetime region under consideration  lies in the past of the null boundary of the WDW patch \cite{Lehner:2016vdi}.

We change integration variables from $(\lambda, \theta)$ to $(\mu, z)$ using the Jacobian determinant 
\beq
J
= \frac{1}{\alpha \, \cos \le \mu_0 - \mu \ri}  \, ,
\label{eq:Jacobian_determinant}
\eeq
leading to the following form:
\beq
I^{\rm WDW}_{\mathcal{N}} = \frac{L}{\pi G} \int_0^{\mu^*(\varepsilon)} d \mu \int_{\delta}^{z_{\rm int}(\mu)} dz  \,  \sqrt{f(\mu)}
\left[ \frac{\sin \le \mu_0 - \mu \ri}{z} + \frac{\cos \le \mu_0 - \mu \ri}{2 z} \frac{f'(\mu)}{f(\mu)} \right]   \, .
\eeq
We put a symmetry factor of 4 to take into account both the region at negative time 
and the analog boundary term associated to the WDW patch anchored at the boundary $\mu=-\mu_0.$
An explicit evaluation gives
\beq
\begin{aligned}
I^{\rm WDW}_{\mathcal{N}} & = \frac{L}{\pi G} \int_0^{\mu^*(\varepsilon)} d\mu \, 
\left[ \sqrt{f(\mu)} \, \sin \le \mu_0 - \mu \ri  +   \frac{f'(\mu)}{2 \sqrt{f(\mu)}}  \cos \le \mu_0 - \mu \ri \right]  \\
& \times  \left[
\log \le \frac{l}{2 \delta} \ri - \log  \le 1+ \sin \le \mu_0 - \mu \ri \ri  \right] \, .
\end{aligned}
\label{eq:null_boundary_term_JAdS_before_exp}
\eeq

\vskip 4mm \noindent
{\bf Entanglement wedge.} 
We use an affine parameterization for the boundaries of EW,
so  eq. (\ref{termine-azione-bordi-null}) vanishes.

\subsubsection{Joint terms}

The typical structure of a codimension-two joint \cite{Lehner:2016vdi} term reads
 \beq
I_{\mathcal{J}_n} =  \frac{\varepsilon_{\mathfrak{a}}}{8 \pi G} \int_{\mathcal{J}_n} d^{d-1} x \, 
 \sqrt{\gamma} \, \mathfrak{a}  \, ,
\eeq
for the case where at least one null boundary is included, as it will always be the case in the following computation.
The expression of $\mathfrak{a}$ will be specified for each case and involve appropriate scalar products of the null normals.
The coefficients $\varepsilon_{\eta}, \varepsilon_{\mathfrak{a}} = \pm 1$ depend on the orientations of the normal one-forms, while $\gamma$ is the determinant of the induced metric along the codimension-two joint.

\vskip 4mm \noindent
{\bf Joint between the cutoff  $z=\delta$ and the WDW patch.}
This joint involves the timelike surface $z=\delta$ and the null boundary of the WDW patch, therefore it reads
\beq
I_{\mathcal{J}}^{\rm \delta, WDW} = \frac{\varepsilon_{\eta}}{2 \pi G}
\int_0^{\mu^* (\varepsilon)} d\mu \, \sqrt{\gamma} \, \log | \mathbf{k}_R \cdot \mathbf{n}_1 |   \, ,
\eeq
where $\mathbf{k}_R$ and $\mathbf{n}_1$ were defined in eq.~\eqref{eq:normal_one_form_WDW_patch_Janus} and \eqref{eq:normal_n1_JAdS}, respectively.

We put a symmetry factor of 2 for the integration along $\mu,$ and another factor of 2 due to the presence of two joints of this kind, at positive and negative times.
We determine the sign $\varepsilon_{\eta}$ according to \cite{Lehner:2016vdi}: with respect to the null boundary of the WDW patch, the joint is a past boundary and the outward direction is a future one.
Therefore we get the sign
\beq
\varepsilon_{\eta} = -1 .
\eeq
The induced metric is determined by imposing that the joint is located at constant $z=\delta,$ and that the intersection between such cutoff surface and the null boundary of the WDW patch is given by
\beq
t_{\rm WDW} (\mu,\delta) =  \delta  \sin \le \mu_0 - \mu \ri \, .
\eeq
In this way we get
\beq
ds^2_{\rm ind} = L^2 f(\mu) \left[ 1 - \frac{1}{z^2} \le \frac{dt_{\rm WDW} (\mu,\delta)}{d\mu} \ri^2  \right] d \mu^2 \, ,
\eeq
with induced metric determinant
\beq
\sqrt{\gamma} = L \sqrt{f(\mu)} \sin \le \mu_0 - \mu \ri \, .
\eeq
The integrand reads
\beq
\log | \mathbf{k}_R \cdot \mathbf{n}_1 | = \log \left|  \frac{\alpha L}{z} \sqrt{f(\mu)} \sin \le \mu_0 - \mu \ri  \right| \, .
\eeq
Putting everything together and evaluating the terms at $z=\delta,$ we find
\beq
I^{\delta, \rm WDW}_{\mathcal{J}} = 
- \frac{L}{2 \pi G} \int_0^{\mu^*(\varepsilon)} d\mu \,  \sqrt{f(\mu)} \, 
\sin \le \mu_0 - \mu \ri \, 
\left[ \log \le  \frac{\alpha L}{\delta} \ri + \log \le  \sqrt{f(\mu)} \sin \le \mu_0 - \mu \ri  \ri \right] \, .
\label{eq:joint1_I_before_exp}
\eeq

\vskip 4mm \noindent
{\bf Joint between the cutoff  $\mu=\mu^*(\varepsilon)$ and the WDW patch.}
We remind that the cutoff surface located at $\mu=\mu^*(\varepsilon)$ intersects both the
 WDW patch and the EW, therefore we need to consider two joint contributions arising from such spacetime region.
The first one corresponds to an expression of the kind
\beq
I_{\mathcal{J}}^{\rm \varepsilon, WDW} = \frac{\varepsilon_{\eta}}{2 \pi G}
\int_{\delta}^{z_{\rm int} (\mu^*(\varepsilon))} d z \, \sqrt{\gamma} \, \log | \mathbf{k}_R \cdot \mathbf{n}_2 |   \, ,
\eeq
where $\mathbf{k}_R$ and $\mathbf{n}_1$ were defined in eq.~\eqref{eq:normal_one_form_WDW_patch_Janus} and \eqref{eq:normal_n2_JAdS}, respectively.

The factor of 4 comes from the symmetry along the time direction and the fact that we include both the joints at $\mu=\pm \mu^*(\varepsilon).$
The joint is a past boundary for the null surface of the WDW patch, and the outward direction points towards the future.
Therefore, we find again
\beq
\varepsilon_{\eta} = -1 \, .
\eeq
The induced metric is
\beq
ds^2_{\rm ind} = \frac{L^2 f(\mu^* (\varepsilon))}{z^2}
\left[ 1 - \le \frac{dt_{\rm WDW} (\mu^* (\varepsilon), z)}{dz} \ri^2   \right] dz^2 \, ,
\eeq
with metric determinant
\beq
\begin{aligned}
\sqrt{\gamma}  =  \frac{L}{z \, \varepsilon}  \cos \le \mu_0 - \mu^* (\varepsilon) \ri   \, ,
\end{aligned}
\eeq
having used the definition of the cutoff in eq.~\eqref{eq:def_cutoff_Janus_AdS}.

We also compute the integrand
\beq
\log | \mathbf{k}_R \cdot \mathbf{n}_2 | = 
\log \left|  \frac{\alpha L}{z} \sqrt{f(\mu)} \cos \le \mu_0 - \mu \ri  \right|_{\mu=\mu^* (\varepsilon)} =
\log \left| \frac{\alpha L}{z \varepsilon}  \cos \le \mu_0 - \mu^*(\varepsilon) \ri \right|  \, .
\eeq
Therefore we obtain
\beq
\begin{aligned}
& I^{\varepsilon, \rm WDW}_{\mathcal{J}}  =   \frac{L}{4 \pi G \, \varepsilon} \cos \le \mu_0 - \mu^*(\varepsilon) \ri 
\left\lbrace
2 \log \left[ \frac{\alpha L}{\varepsilon} \cos \le \mu_0-\mu^*(\varepsilon) \ri   \right] \log \le \frac{2 \delta}{l} \ri  \right. \\
& \left. + 2 \log \left[ \frac{\alpha L}{\varepsilon} \cos \le \mu_0-\mu^*(\varepsilon) \ri   \right] \log \left[ 1+ \sin \le \mu_0 - \mu^*(\varepsilon) \ri \right]
- \log^2 \delta + \log^2 \left[ \frac{2 \le 1+ \sin \le \mu_0 - \mu^*(\varepsilon) \ri  \ri}{l}  \right]
\right\rbrace  \, .
\end{aligned}
\label{eq:joint_term_NI}
\eeq

\vskip 4mm \noindent
{\bf Joint between the cutoff $\mu=\mu^*(\varepsilon)$ and the EW.}
The computation is similar to the previous joint, but in this case the intersection involves the EW instead of the WDW patch.
The term to include in the action is 
\beq
I^{\varepsilon, \rm EW}_{\mathcal{J}} = 
\frac{\varepsilon_{\eta}}{2 \pi G} \int_{z_{\rm int} (\mu^*)}^{z_{\rm RT}} dz \, \sqrt{\gamma} \, \log | \mathbf{w} \cdot \mathbf{n}_2 | \, ,
\eeq
where $\mathbf{w}$ and $\mathbf{n}_2$ were defined in eq.~\eqref{eq:null_one_form_EW} and \eqref{eq:normal_n2_JAdS}, respectively.

The factor of 4 comes from the fact that there are two joints of this kind, for both positive and negative times, and there is another joint located at $\mu=-\mu^*(\varepsilon).$
The induced metric in this case is
\beq
ds^2_{\rm ind} = \frac{L^2 f(\mu)}{z^2} \left[ 1 - \le  \frac{dt_{\rm EW} (z)}{dz} \ri ^2 \right] dz^2 = 0 \, ,
\eeq
and the scalar product between the normals is
$ \mathbf{w} \cdot \mathbf{n}_2  = 0 $.
The result seems divergent because the last term enters as the argument of a logarithm,
 but since the induced metric determinant vanishes as well and faster, we  find
$I^{\varepsilon, \rm EW}_{\mathcal{J}} =  0$.

\vskip 4mm \noindent
{\bf Joint on the RT surface.}
There is a joint precisely at the RT surface located at $t=0$ and $z=l/2,$ arising
from the intersection between the positive and negative parts of the entanglement wedge.
The structure of the integration is given by
\beq
I_{\mathcal{J}}^{\rm RT} = \frac{\varepsilon_{\eta}}{4 \pi G} 
\int_0^{\mu^*(\varepsilon)} d\mu \, 
\sqrt{\gamma} \log \left| \frac{1}{2} \mathbf{w}_+ \cdot \mathbf{w}_- \right|  \, , 
\eeq
where the symmetry factor of 2 comes from the integration along $\mu.$
Here we denoted with $\mathbf{w}_{\pm}$
 the normals to the positive and negative parts of the entanglement wedges, respectively
\beq
 \mathbf{w}_{\pm}= \b(\pm dt  +dz) \, .
\eeq

The RT surface  represents for the upper (lower) portion of the entanglement wedge a past (future) boundary, and the outward normal points towards the future (past).
Therefore, the sign is given by
\beq
\varepsilon_{\eta} = -1 \, .
\eeq
The induced metric is
\beq
ds^2_{\rm ind} = L^2 f(\mu) d\mu^2 \, ,
\eeq
since the RT surface sits at constant $t=0$ and $z=l/2.$
Therefore the metric determinant is
\beq
\sqrt{\gamma} = L \sqrt{f(\mu)} \, . 
\eeq
The scalar product between the null normals reads
\beq
\frac{1}{2} \, \mathbf{w}_+ \cdot \mathbf{w}_- =  \frac{\beta^2}{L^2} \frac{z^2}{f(\mu)} \, .
\eeq
Collecting these factors together, we obtain
\beq
I_{\mathcal{J}}^{\rm RT}
=- \frac{L}{2 \pi G} 
\int_0^{\mu^*(\varepsilon)} d\mu \, \sqrt{f(\mu)} \, 
\log \left| \frac{\beta}{2 L} \frac{l}{\sqrt{f(\mu)}} \right|  \, .
\label{eq:joint2_I_before_exp}
\eeq

\vskip 4mm \noindent
{\bf Joint at the intersection  between WDW patch and EW.}
There are a couple of joints (located at positive or negative times) at the intersection curve between the WDW patch and the entanglement wedge.
By symmetry, we compute only one of them and we multiply the result by a factor of 2.
The structure is the following:
\beq
I_{\mathcal{J}}^{\rm int} = \frac{\varepsilon_{\eta}}{2 \pi G} 
\int_0^{\mu^* (\varepsilon)} d \mu \, \sqrt{\gamma} 
\log \left| \frac{1}{2} \mathbf{w} \cdot \mathbf{k}_R \right| \, ,
\eeq
where another factor of 2 arises due to the integration over $\mu.$
The sign is given by
\beq
\varepsilon_{\eta} = 1 \, ,
\eeq
since for both null surfaces the joint represents the future boundary and the outward-directed normal also points to the future.
The induced metric is
\beq
ds^2_{\rm ind} = L^2 f(\mu) \left[  1 + \frac{1}{z^2} \le \frac{d z_{\rm int}(\mu)}{d\mu} \ri^2 - \frac{1}{z^2} \le \frac{d t_{\rm int} (\mu)}{d\mu} \ri^2  \right] d\mu^2 \, ,
\eeq
where in the last part we express the intersection curve as a function $t_{\rm int} (\mu)$ by plugging in the function \eqref{eq:intersection_curve_Janus_AdS} inside either $t_{\rm WDW} (\mu,z)$ or $t_{\rm EW}(z).$
Working in this way, we obtain
\beq
t_{\rm int}(\mu) = \frac{l}{2} \le 1- \frac{1}{\sin \le \mu_0 - \mu \ri +1} \ri \, .
\eeq
The metric determinant simplifies to
\beq
\sqrt{\gamma} = L \sqrt{f(\mu)} \, ,
\eeq
and the argument of the logarithmic term in the integral reads
\beq
\frac{1}{2} \, \mathbf{k}_R \cdot \mathbf{w} = 
- \frac{\alpha \beta}{2} \left[ 1+ \sin \le \mu_0 - \mu \ri  \right] \, .
\eeq
Therefore, the corresponding joint term is given by
\beq
I_{\mathcal{J}}^{\rm int} =  \frac{L}{2 \pi G} 
\int_0^{\mu^*(\varepsilon)} d \mu \, 
\sqrt{f(\mu)} \, 
\log \left| \frac{\alpha \beta}{2} \left[ 1+ \sin \le \mu_0 - \mu \ri  \right] \right| \, .
\label{eq:joint3_I_before_exp}
\eeq

\vskip 4mm \noindent
{\bf Joint  at $\mu=0$.} The last kind of joint term arises from the intersection between the WDW patches at $\mu=0.$
By symmetry, there are two of them, at positive and negative values of time.
The integral reads
\beq
I_{\mathcal{J}}^{\mu} = \frac{\varepsilon_{\eta}}{4 \pi G}
\int_{\delta}^{z_{\rm int} (0)} dz \, \sqrt{\gamma} \, \log \left| \frac{1}{2} \mathbf{k}_L \cdot \mathbf{k}_R  \right| \, ,
\eeq
where $\varepsilon_{\eta} = -1$ and
 $\mathbf{k}_{L,R}$ denotes the null normals to the WDW patch on the left and right sides of the surface $\mu=0$,
see eq.  \eqref{eq:normal_one_form_WDW_patch_Janus}.

We compute the induced metric
\beq
ds^2_{\rm ind} = \frac{L^2 f(\mu)}{z^2} \left[ 1 - \le \frac{dt_{\rm WDW} (0,z)}{dz} \ri^2  \right] dz^2 \, ,
\eeq
and the metric determinant
\beq
\sqrt{\gamma}  =
- \frac{\alpha_+ L}{z} \cos \mu_0
  \, ,
\eeq
where in the last step we used that $f(0)= \alpha_+^2 .$
The scalar product between the null one-forms reads
\beq
\left| \frac{1}{2} \, \mathbf{k}_L \cdot \mathbf{k}_R \right| =
 \frac{\alpha^2 L^2}{z^2} \, \alpha_+^2 \cos^2 \mu_0 \, .
\eeq
Since the integral involves only the $z$ variable, it can be explicitly evaluated:
\beq
\begin{aligned}
I^{\mu}_{\mathcal{J}} &  = - \frac{L \alpha_+ \, \cos \mu_0}{4 \pi G}
\left\lbrace  
2 \log \left| \alpha L \alpha_+ \cos \mu_0  \right|
\log \le \frac{2 \delta}{l} \ri  \right. \\
& \left. + 2 \log | \alpha L \alpha_+ \cos \mu_0 |
\log \le 1+ \sin \mu_0 \ri
- \log^2 \delta 
+ \log^2 \le \frac{2(1+ \sin \mu_0)}{l} \ri
 \right\rbrace   \, .
\end{aligned}
\label{eq:joint_term2_NI}
\eeq

\subsubsection{Counterterm on null boundaries}

Since the gravitational action is not reparametrization-invariant as it stands, we need to add a counterterm 
 \cite{Lehner:2016vdi} on null boundaries which reads
\beq
I_{\rm ct} = \frac{1}{8 \pi G} \int_{\mathcal{B}_n} 
d\lambda d^{d-1} x \, \sqrt{\gamma} \,  \Theta \, \log |\tilde{L} \Theta| \, ,
\label{termine-azione-controtermine}
\eeq 
where $\gamma$ is the induced metric determinant along the coordinates orthogonal to the parameter $\lambda,$ 
 $\Theta$ is the expansion of the congruence of null geodesics and $\tilde{L}$ is an arbitrary scale.

\vskip 4mm \noindent
{\bf WDW patch.}
Using the parametrization \eqref{eq:parametrization_WDW_patch_Janus_AdS}
 for the null boundary of the WDW patch and the corresponding expansion determined in
 eq.~\eqref{gamma-theta-kappa-left},  we obtain
\beq
\begin{aligned}
I_{\rm ct}^{\rm WDW} & = \frac{L}{2 \pi G} \int_0^{\mu^*(\varepsilon)}
d \mu \, \int_{\delta}^{z_{\rm int}(\mu)} d z \,  
\left[ \frac{\sqrt{f(\mu)} \sin \le \mu_0 - \mu \ri}{z} + \frac{\cos \le \mu_0 - \mu \ri}{2 z} \frac{f'(\mu)}{\sqrt{f(\mu)}} 
\right]  \\
& \times \log \left| \alpha \tilde{L} \left[ \frac{\sin \le \mu_0 - \mu \ri}{z} + \frac{\cos \le \mu_0 - \mu \ri}{2 z} \frac{f'(\mu)}{f(\mu)} 
\right]   \right|
\end{aligned}
\eeq
where we put a symmetry factor of 4 (due to the time and $\mu$ coordinates) and 
we used the Jacobian determinant \eqref{eq:Jacobian_determinant} to change variables.
A direct calculation gives
\beq
\begin{aligned}
I_{\rm ct}^{\rm WDW} & = \frac{L}{4 \pi G} \int_0^{\mu^*(\varepsilon)}
d \mu \, 
\left[ \sqrt{f(\mu)} \sin \le \mu_0 - \mu \ri + \cos \le \mu_0 - \mu \ri \frac{f'(\mu)}{2 \sqrt{f(\mu)}} 
\right] \\
& \times \left\lbrace \log^2 \left[ \frac{\alpha \tilde{L}}{\delta} \, \le \sin \le \mu_0 - \mu \ri + \cos \le \mu_0 - \mu \ri \frac{f'(\mu)}{2 f(\mu)}  \ri  \right]   \right. \\
& \left. -  \log^2 \left[ \frac{2 \alpha \tilde{L}}{l} \, \le \sin \le \mu_0 - \mu \ri + \cos \le \mu_0 - \mu \ri \frac{f'(\mu)}{2 f(\mu)}  \ri  \le 1 + \sin \le \mu_0 - \mu \ri  \ri  \right]
 \right\rbrace  \, .
\end{aligned}
\label{eq:counterterm_JAdS_before_exp}
\eeq

\vskip 4mm \noindent
{\bf Entanglement wedge.}
The expansion parameter $\Theta$ vanishes on the EW and so the term
in eq. (\ref{termine-azione-controtermine}) vanishes.

\subsubsection{Series expansion of the gravitational action}
\label{sect-technique_series_expansion_action_JAdS}

In the previous subsections we computed term by term the 
contributions to the action in Janus AdS$_3$ spacetime. 
It reads
\beq
I_{\rm tot} (\gamma) = \sum_{\mathcal{X}}  I_{\mathcal{X}} (\gamma) \, ,
\label{eq:tot_grav_action_JAdS_pre_subtr}
\eeq
where the subscript $\mathcal{X}$  runs over 
\beq
\mathcal{X} \in \lbrace \mathcal{B}, \rm GHY, \mathcal{N}, \mathcal{J} ,\rm ct  \rbrace \, ,
\label{eq:listX}
\eeq
\emph{i.e.} it contains the terms defined in eq.~\eqref{eq:total_gravitational_action}: bulk, Gibbons-Hawking-York, null codimension-one boundaries, joints and counterterm.

A generic contribution $I_{\mathcal{X}}$ to the gravitational action is of two kinds:
\begin{enumerate}
\item All the integrations have been performed explicitly. This is the case 
of eqs.~\eqref{eq:GHY_term2_JAdS_before_exp}, \eqref{eq:joint_term_NI}, \eqref{eq:joint_term2_NI}.
In this case, we simply perform a Laurent expansion around $\varepsilon=0$
 as explained at the beginning of Appendix \ref{app-details_expansion_Janus_AdS}.
\item There is still a remaining integration over $\mu$ to perform. This is the case of eqs.~\eqref{eq:bulk_term_JAdS_before_exp}, \eqref{eq:GHY_term1_JAdS_before_exp}, \eqref{eq:null_boundary_term_JAdS_before_exp}, \eqref{eq:joint1_I_before_exp}, \eqref{eq:joint2_I_before_exp}, \eqref{eq:joint3_I_before_exp}, \eqref{eq:counterterm_JAdS_before_exp}.
In this case, we need to extract the singular behaviour as explained below.
\end{enumerate}

The steps that we perform in case 2 are the following:
\begin{itemize}
\item Consider a generic integral among the list of case 2, \emph{i.e.}
\beq
I (\mu_0) = \int_{0}^{\mu^*(\varepsilon)} d\mu \, \mathcal{F} (\mu) \, ,
\label{eq:typical_integral_to_evaluate}
\eeq
where $\mu^*(\varepsilon)$ is the regulator defined in eq.~\eqref{eq:def_cutoff_Janus_AdS} such that it collapses to $\mu_0$ when $\varepsilon=0,$ but is otherwise chosen to avoid that $\mu_0$ enters the integration domain.
If the integrand $\mathcal{F}$ is singular in $\mu=\mu_0,$ we  Laurent expand it  around $\mu_0.$
\item Collect the terms in the series expansion of $\mathcal{F}$ as follows: 
\beq
\mathcal{F} (\mu) = \mathcal{F}_S (\mu,\mu_0) + \mathcal{F}_R (\mu,\mu_0) \, ,
\label{eq:expansion_calF}
\eeq
where $\mathcal{F}_S$ are singular terms in $\mu_0$ after the integration over $\mu,$ and $\mathcal{F}_R$ are instead regular (\emph{i.e.} analytic).
The number of singular terms will be finite, while the regular part will contain, in principle, infinite terms with positive powers of $(\mu_0-\mu).$
\item Sum and subtract the singular part from the original integral, getting
\beq
I (\mu_0) = \int_{0}^{\mu_0} d\mu \, \le  \mathcal{F} (\mu) - \mathcal{F}_S (\mu,\mu_0) \ri  + 
\int_{0}^{\mu^*(\varepsilon)} d\mu \,  \mathcal{F}_S (\mu)  \, .
\label{eq:technique_renormalization_action}
\eeq
We analytically compute the last term, since it is now a sum of rational functions; instead we numerically evaluate the former term. 
The advantage of this splitting is that the first part of the solution is regular in $\mu=\mu_0,$ and therefore we can evaluate the limit $\varepsilon \rightarrow 0$ explicitly.
\end{itemize}
The result of this method is a bunch of numerical functions analytic in $\mu_0,$ plus divergent terms evaluated with this regularization procedure.
After performing the integrals along $\mu$ in this way, we expand the remaining terms around $\varepsilon=0.$ 
The details of the computation are explained term by term in Appendix \ref{sect-app-expansion_term_by_term}.
At the end of the procedure, the gravitational action is given by eq.~\eqref{eq:tot_grav_action_JAdS_pre_subtr}, where the five terms entering the set \eqref{eq:listX} are written in eqs.~\eqref{eq:bulk_term_expanded}, \eqref{eq:GHY_expanded}. \eqref{eq:I_N_expanded}, \eqref{eq:IJ_expanded}, \eqref{eq:Ict_expanded}.

\subsubsection{Subtraction of the empty AdS solution and final result}

The double cutoff regularization requires to subtract the subregion action evaluated in vacuum AdS$_3$ 
(which is recovered in the $\gamma \rightarrow 0$ limit)
and verify that the quantity obtained in this way is independent of the parameter $\varepsilon$ introduced above. 
This step can be performed using the following results:
\beq
\begin{aligned}
& \lim_{\gamma \rightarrow 0} \alpha_+ = 1 \, , \quad
  \lim_{\gamma \rightarrow 0} \alpha_- = 0 \, , \quad
  \lim_{\gamma \rightarrow 0} \mu_0 = \frac{\pi}{2} \, , \quad
  \lim_{\gamma \rightarrow 0} f(\mu) = \frac{1}{\cos^2 \mu} \, , \quad
\lim_{\gamma \rightarrow 0} \mu^*(\varepsilon) = 
\frac{\pi}{2} - \arcsin \varepsilon \, .
\end{aligned}
\label{eq:useful_limits_gamma_to_zero}
\eeq
In this way, we get the contribution to the subregion complexity  intrinsic to the defect 
\beq
\begin{aligned}
\Delta \mathcal{C}_A^{\rm JAdS} (\gamma) = & 
\frac{L}{\pi^2 G} \,  P \left( \gamma, {\tilde{L}}/{ L} \right) \,
 \log \le \frac{l}{\delta} \ri + \, \, {\rm finite \,\, terms}  \, ,
\end{aligned} 
\label{eq:final_divergence_action}
\eeq
where
\beq
\begin{aligned}
P \left( \gamma, {\tilde{L}}/{ L} \right) & = - \mathcal{I}_{\mathcal{B}}^{(1)} (\gamma) 
 + \frac{1}{2} \le  \frac{\pi}{2} -  \mathcal{I}_{\rm GHY} (\gamma) \ri 
- \frac{1}{4} \le \mathcal{I}_{\rm ct}^{(1)} (\gamma) - \mathcal{I}_{\rm ct}^{(1)} (0) \ri  \\
& + \frac{3}{2 \mu_0} - \frac{1}{\pi}
- \alpha_+ \cos \mu_0 \le 1 + \frac{1}{2} \log \left| \frac{\tilde{L}}{L} \frac{\tan \mu_0}{\alpha_+} \right| \ri \, .
 \end{aligned}
 \label{eq:prefactor_log_divergence}
\eeq
The functions $ \mathcal{I}_{\mathcal{B}}^{(1)} $, $ \mathcal{I}_{\rm GHY}$ and  $\mathcal{I}_{\rm ct}^{(1)}$
 are defined in eqs.~\eqref{eq:numerical_function_IB1}, \eqref{eq:numerical_function_IGHY},
  \eqref{eq:numerical_function_Ict1}, while $\a_+$ and $\mu_0$
  were introduced in eq.~\eqref{eq:data_elliptic_functions}.

We observe that the contribution intrinsic to the defect  $\Delta \mathcal{C}_A^{\rm JAdS}$ is 
logarithmically divergent in the ratio between  the length $l$ of the subregion on the boundary and the UV cutoff $\delta.$
All the dependence in the computation from the ambiguity in normalizing the null normals, parametrized by $\alpha$ and $\b$, disappear in the final result.
On the contrary, $\Delta \mathcal{C}_A^{\rm JAdS}$ depends on the arbitrary scale $\tilde{L}$
which enters in the counterterm in eq.~(\ref{termine-azione-controtermine}).

\begin{figure}[ht]
\centering
\includegraphics[scale=1.2]{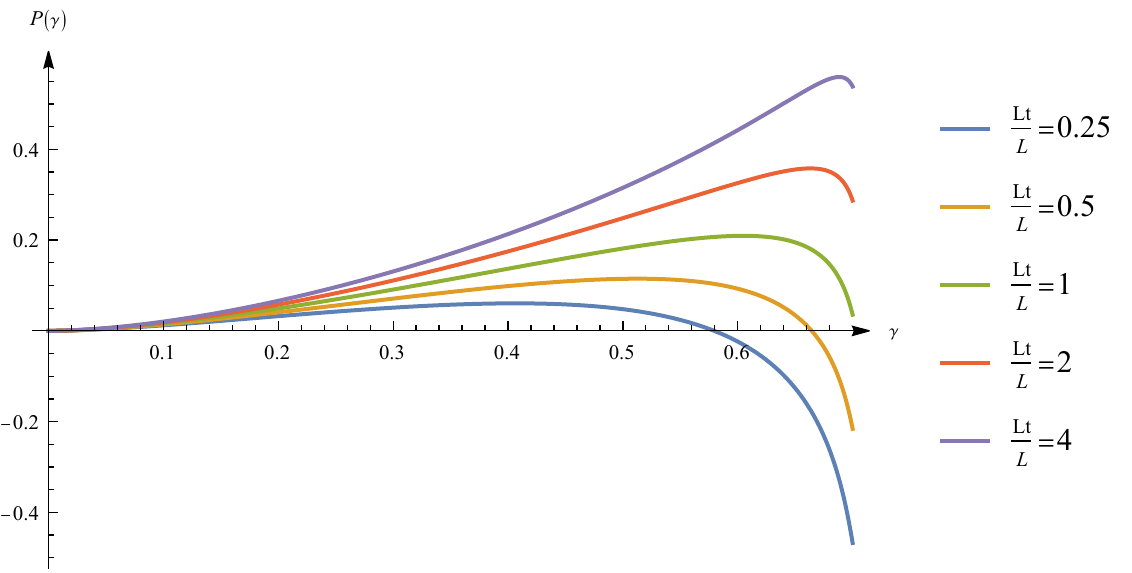}
\caption{ Plot of the expression defined in eq.~\eqref{eq:prefactor_log_divergence} 
as a function of $\gamma.$ Various colors correspond to different choices
 of  $\tilde{L}/L$.}
\label{fig-coeff_divergence}
\end{figure}

According to the discussion in \cite{Baiguera:2021cba}, we expect that the coefficient $ P ( \gamma, {\tilde{L}}/{ L} )$ of the logarithmic divergence
is  independent of the regularization prescription.
The quantity $ P ( \gamma, {\tilde{L}}/{ L} )$
 is numerically evaluated in Fig.~\ref{fig-coeff_divergence} as a function of $\gamma$ for various choices of $\tilde{L}/L$.
Note that, for small deformation parameter  $\gamma$, the quantity $\Delta \mathcal{C}_A^{\rm JAdS}$
is positive, meaning  that it is computationally harder to produce an interface than vacuum space.
Since our calculation is only valid for $\gamma \leq \gamma_0 \approx 0.704$, see eq. (\ref{gamma-zero-assumption}),
we can not give any physical meaning to the divergence of $P$ for $\gamma \to 1 / \sqrt{2} \approx 0.707$.

\begin{figure}[ht]
\begin{center}
 \def\svgwidth{\columnwidth}
    \scalebox{0.6}{
\begingroup%
  \makeatletter%
  \providecommand\color[2][]{%
    \errmessage{(Inkscape) Color is used for the text in Inkscape, but the package 'color.sty' is not loaded}%
    \renewcommand\color[2][]{}%
  }%
  \providecommand\transparent[1]{%
    \errmessage{(Inkscape) Transparency is used (non-zero) for the text in Inkscape, but the package 'transparent.sty' is not loaded}%
    \renewcommand\transparent[1]{}%
  }%
  \providecommand\rotatebox[2]{#2}%
  \newcommand*\fsize{\dimexpr\f@size pt\relax}%
  \newcommand*\lineheight[1]{\fontsize{\fsize}{#1\fsize}\selectfont}%
  \ifx\svgwidth\undefined%
    \setlength{\unitlength}{749bp}%
    \ifx\svgscale\undefined%
      \relax%
    \else%
      \setlength{\unitlength}{\unitlength * \real{\svgscale}}%
    \fi%
  \else%
    \setlength{\unitlength}{\svgwidth}%
  \fi%
  \global\let\svgwidth\undefined%
  \global\let\svgscale\undefined%
  \makeatother%
  \begin{picture}(1,0.65020027)%
    \lineheight{1}%
    \setlength\tabcolsep{0pt}%
    \put(0,0){\includegraphics[width=\unitlength,page=1]{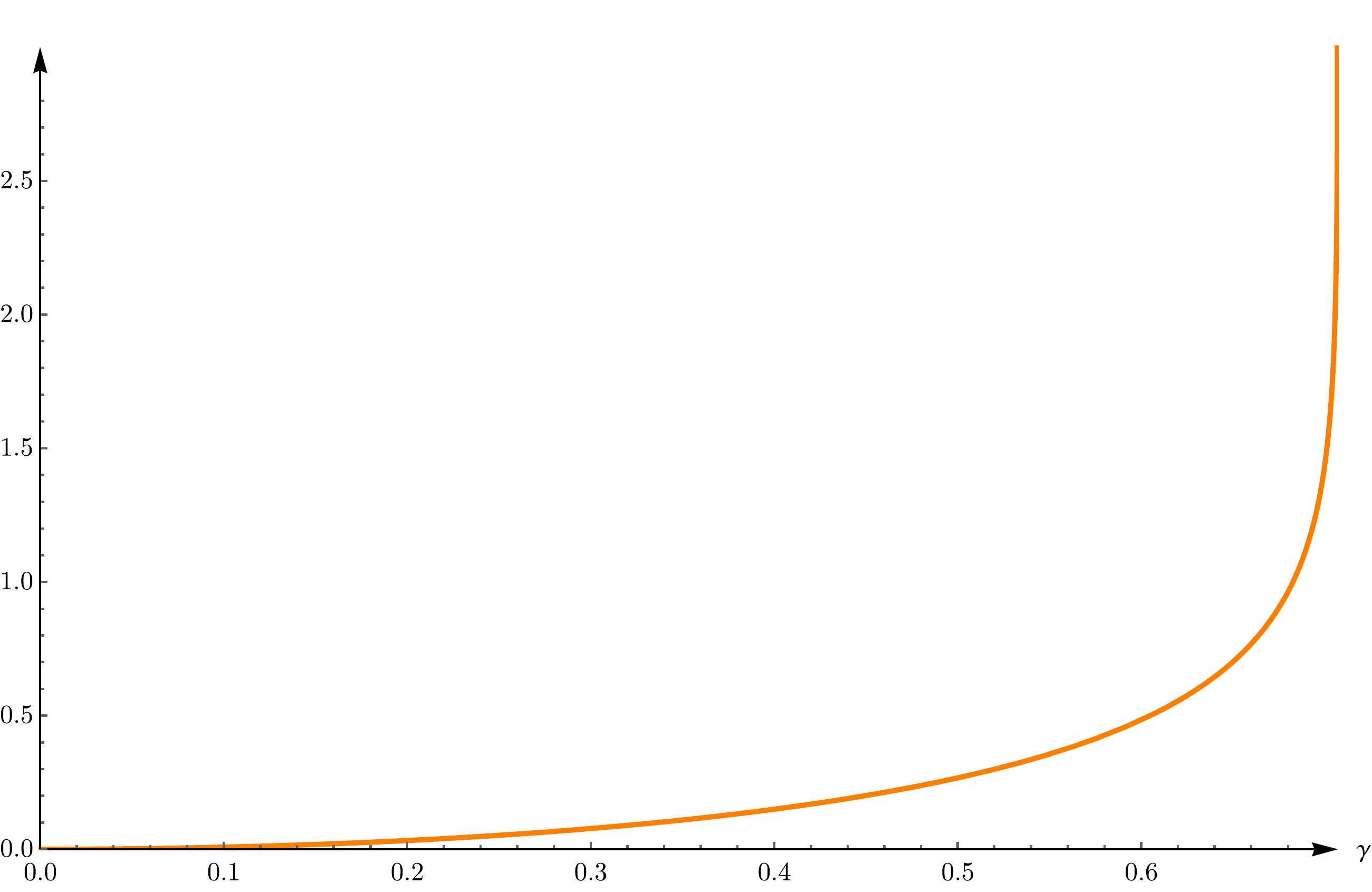}}%
    \put(0.01096723,0.62591568){\makebox(0,0)[lt]{\lineheight{1.25}\smash{\begin{tabular}[t]{l}$\eta(\gamma)$\end{tabular}}}}%
    \put(0,0){\includegraphics[width=\unitlength,page=2]{eta_di_gamma.pdf}}%
    \put(0.97765045,0.02271783){\makebox(0,0)[lt]{\lineheight{1.25}\smash{\begin{tabular}[t]{l}$\gamma$\end{tabular}}}}%
  \end{picture}%
\endgroup%
}
\caption{
Plot of $\eta(\gamma)$ as defined in Eq.~\eqref{eq:eta_gamma}, which is the coefficient
of the log divergences due to the defect in the volume case. 
}
\label{eta-gamma}
\end{center}
\end{figure}

The divergence structure obtained in Eq.~\eqref{eq:final_divergence_action} is of the same kind as the one predicted using the volume conjecture \cite{Auzzi:2021nrj}.
Indeed, we found using the double cutoff regularization:
\beq
\Delta \mathcal{C}_V^{\rm JAdS} =  \frac{L}{G} \, \eta(\gamma)\log\left(\frac{l}{ \delta} \right)
 + \, \, {\rm finite \,\, terms}  \,,
\label{eq:volume_Janus_AdS}
\eeq
where 
\beq
\eta(\gamma)= 2 \alpha_+ \le \mathbb{K}(m)- \mathbb{E}(m)  \ri  \, .
\label{eq:eta_gamma}
\eeq
The plot of the function $\eta(\gamma)$ is shown in Fig.~\ref{eta-gamma}.
The contribution intrinsic to the defect  $\Delta \mathcal{C}_V^{\rm JAdS}$
 is always  positive and diverges when $\gamma \rightarrow 1/\sqrt{2}$.

 At small $\gamma$, the coefficient of the volume divergent term \cite{Auzzi:2021nrj}
scales as $\gamma^2$, i.e. $\eta \approx \pi \gamma^2/4 $.
From the eq. (\ref{eq:prefactor_log_divergence}),
we checked that the coefficient $\mathcal{O}(\gamma)$ of the small $\gamma$ expansion of $P$ indeed vanishes, i.e.
\beq
P ( \gamma, {\tilde{L}}/{ L} )= c_1 \gamma^2  + c_2 \gamma^2 \log \gamma^2 + \mathcal{O}(\gamma^3) \, ,
\eeq
where 
\beq
c_1 = \frac{15}{8 \pi}  - \frac{5 \pi}{64}
 + \frac{3 \pi}{32} \log \left| \frac{8}{3 \pi} \frac{\tilde{L}}{L} \right| \, , \qquad
c_2 = - \frac{3 \pi}{32} \, .
\eeq
The divergent parts of volume and action have then a similar 
(even if not identical) parametric dependence
at small $\gamma$.

\section{Subregion complexity in the AdS$_3$/BCFT$_2$ model}
\label{sect-computation_action_BCFT}

In this section we compute the subregion complexity in the AdS$_3$/BCFT$_2$ model.
The CFT is restricted to live on a half plane of the flat spacetime $x \geq 0$, because
there is a boundary at $x=0$.
In the present work we are interested to determine if the presence of 
this boundary
 entails a logarithmically divergent complexity.
In principle, we could  consider the case of an arbitrary interval  $x \in [l_1, l_2]$,
which, in general, does not contain the boundary at $x=0$.
If the interval does not contain the boundary,
we do not expect extra divergences in addition to the ones of pure AdS
(this was explicitly checked for the CV case in \cite{Braccia:2019xxi}).
 Therefore we   will consider the subregion complexity of the interval $x \in [0, l/2] .$ 

In section \ref{sect-null_boundaries_BCFT} we specify the domain of integration,
see Fig.~\ref{fig-geometrical_setting_BCFT}.
 We perform the calculation of the action in section
 \ref{sect-computation_action_BCFT-action-terms}.
 In order to find the intrinsic contribution coming from the presence of the brane, we will subtract
 the vacuum solution.

\subsection{Null boundaries in the AdS$_3$/BCFT$_2$ model}
\label{sect-null_boundaries_BCFT}

We  work with the metric in Poincaré coordinates $X^\mu=(t,z,x)$, see eq.~\eqref{eq:metric_AdS_BCFT}.
The bulk dual geometry is delimited by the end-of-the-world brane
 \beq
 x = - z \, \cot \alpha \, .
 \label{posizione-brana}
 \eeq
The RT surface at $t=0$ is  the same as  in pure AdS
\beq
z_{\rm RT} (x)= \sqrt{\le \frac{l}{2} \ri^2 - x^2} \, .
\label{RT-ads-bcft}
\eeq
The EW and most of the WDW patch are also the same as the ones in empty AdS space.
An extra portion  of WDW patch (see \cite{Braccia:2019xxi,Sato:2019kik})  is also needed.

We use the regularization  A in Fig.~\ref{fig-2regs} where the WDW patch and
 the EW start from the true boundary located at $z=0.$ 
The cutoff is the surface $z=\delta.$
At the end of the computation, we will send $\delta \rightarrow 0.$ 

\vskip 4mm \noindent
{\bf WDW patch.} 
Here we consider just the $t>0$ boundary of the WDW,
the $t<0$ part can be found by symmetry.
The null boundary of the WDW patch in the right region $x \geq 0$ originating from the surface at $t=0$ and $z=0$ is
\beq
t_{\rm WDW,R } (z,x) =z \, .
\eeq
In the left region $x<0,$ the boundary of the WDW patch 
is a portion of the cone
\beq
t_{\rm WDW,L } (z,x) =\sqrt{x^2+z^2} \, ,
\eeq
which intersects the brane defined by eq.~(\ref{posizione-brana}).
The null boundary of this portion of WDW patch can 
be parametrized by the congruence of geodesics
\beq
X^{\mu}_{\rm WDW,L} = B \, (\lambda, \lambda \cos \theta, -\lambda \sin \theta) \, ,
\eeq
where each value of $\theta \in [0, \frac{\pi}{2} - \alpha]$ gives a different null geodesic,
 $\l$ is the (non-affine) geodesic parameter and
 $B>0$ is an arbitrary constant.
The relevant geometric quantities are
\beq
k^{\mu}_{\rm WDW,L} = B \, (1, \cos \theta, - \sin \theta) \, , \quad
\kappa_{\rm WDW,L} = - \frac{2}{\lambda} \, , \quad
\sqrt{\gamma_{\rm WDW,L}} = \frac{L}{\cos \theta} \, , \quad
\Theta_{\rm WDW,L} = 0 \, .
\eeq

%

\vskip 4mm \noindent
{\bf Entanglement wedge.}
The boundary of the EW is  the same as  in pure AdS
\beq
t_{\rm EW}(z,x) = \frac{l}{2} - \sqrt{x^2 + z^2} \, .
\label{eq:surface_EW_AdS_BCFT}
\eeq
The tangent vector of the null affine geodesics is
\beq
w^{\mu} = C \, \frac{z^2}{L^2} \le -1, \frac{z}{\sqrt{x^2 +z^2}}, \frac{x}{\sqrt{x^2 + z^2}} \ri \, ,
\label{eq:null_vector_EW_AdS_BCFT}
\eeq
where $C>0$ is a constant.
The expansion parameter $\Theta$ vanishes  
as expected \cite{Headrick:2014cta}.

\begin{figure}[H]
\centering
\includegraphics[scale=0.35]{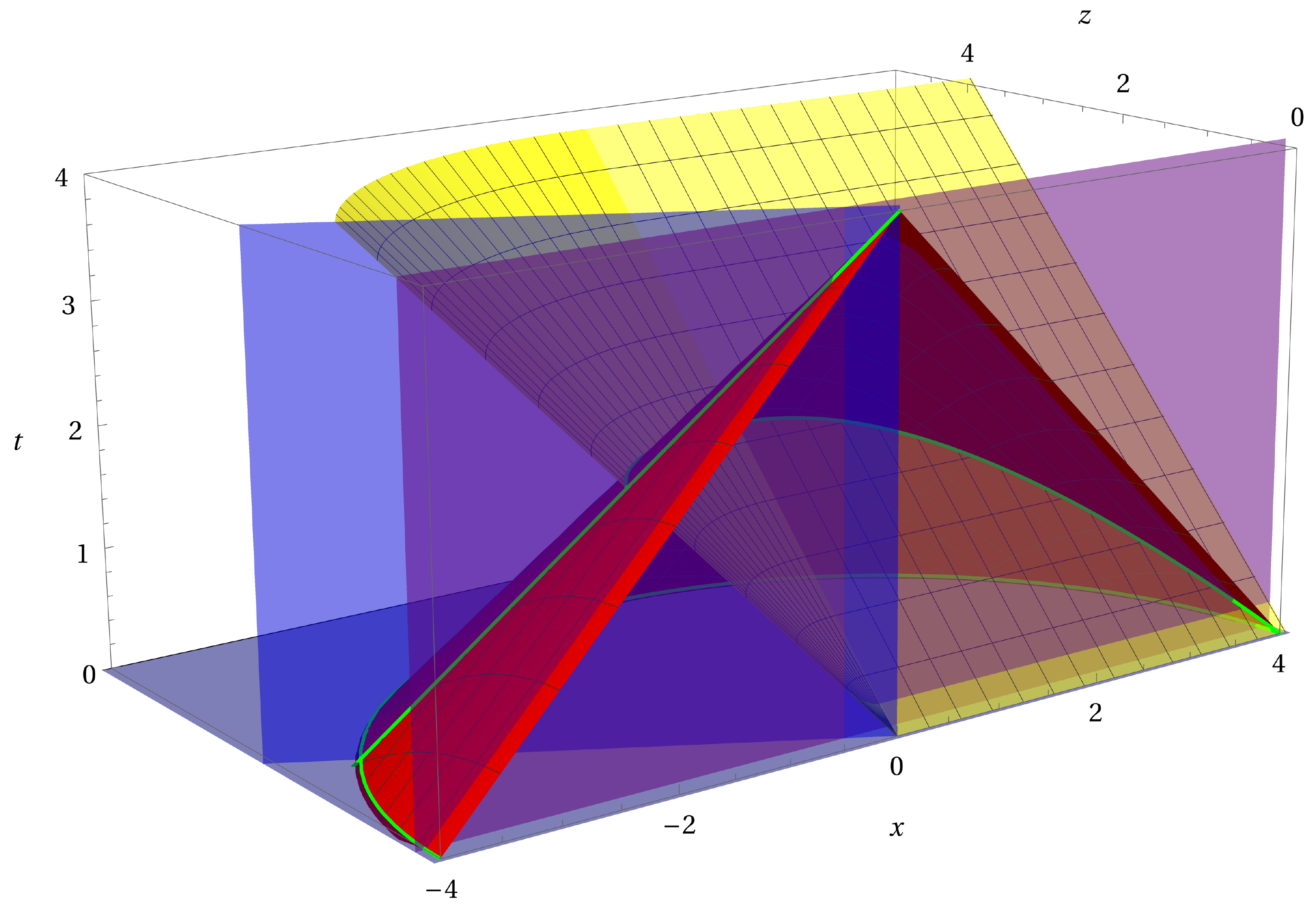}
\includegraphics[scale=0.45]{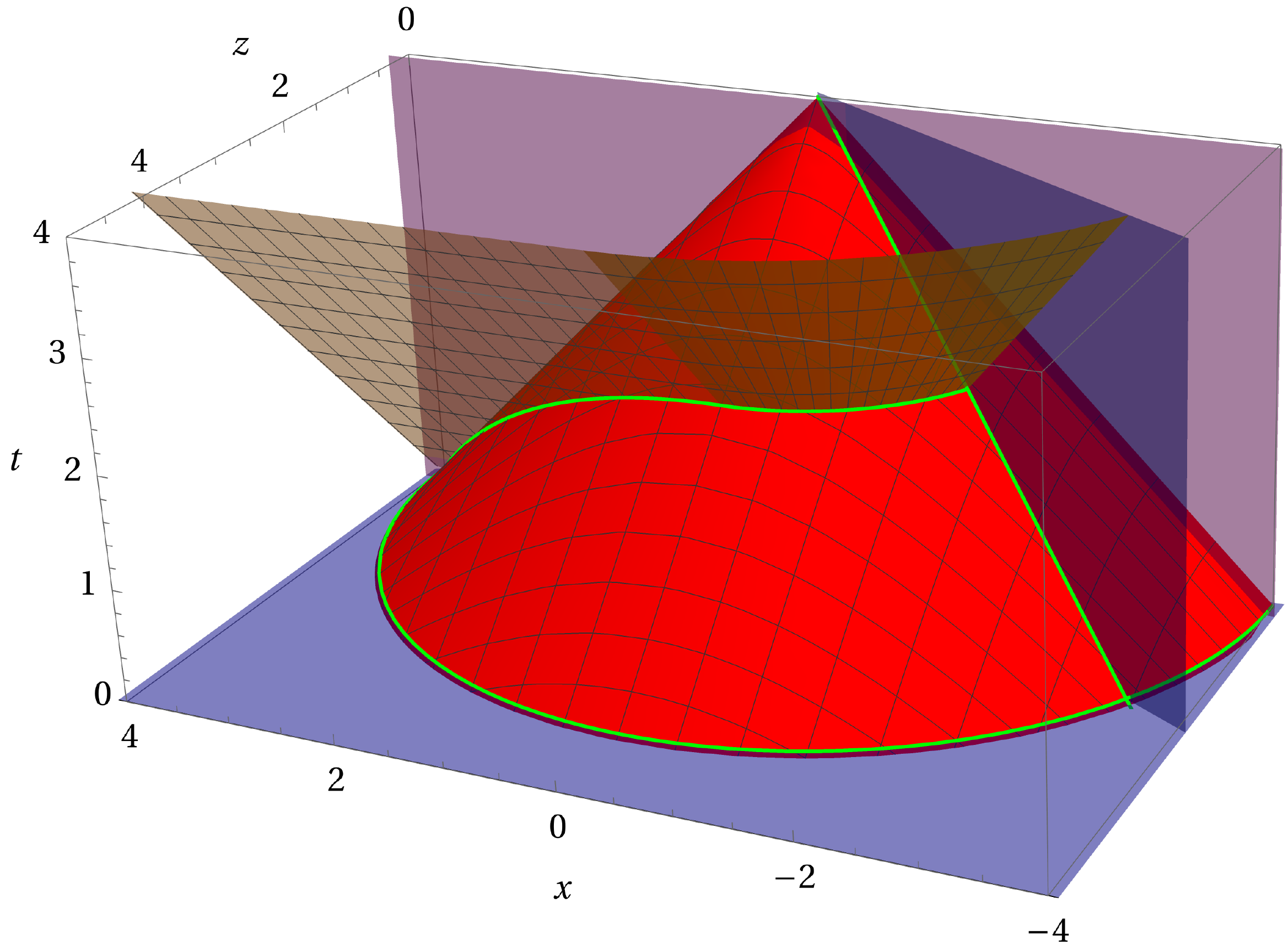}
\caption{Geometrical sketch of the null boundaries of the WDW patch (light yellow surfaces) and of the entanglement wedge (red), taken from two different perspectives.
The blue plane represents the brane and restricts the intergration to the region $x \geq - z \cot \alpha .$
The purple transparent region represents the cutoff surfaces located at $z=\delta.$ 
There are several curves depicted in green representing the intersections between the WDW patch, the entanglement wedge and the brane.
In the figure we take $\alpha= \pi/6 .$}
\label{fig-geometrical_setting_BCFT}
\end{figure}

\vskip 4mm \noindent
{ \bf Intersection between surfaces.}
The following intersection curves play an important role in delimiting the integration regions:
\begin{itemize}
\item The intersection curve between the right side of the WDW patch and the EW:
\beq
x_{\rm int,R} (z)= \frac{1}{2} \sqrt{l(l-4z)} \, ,
\qquad 
 t_{\rm int,R}(z) = z \, .
\label{riccioletto-2}
\eeq
\item The intersection between the left part of the WDW patch and the EW:
\beq
x_{\rm int,L} (z)=  - \sqrt{\le \frac{l}{4} \ri^2 - z^2} \, , \qquad
t_{\rm int,L}(z) = \frac{l}{4} \, .
\label{riccioletto}
\eeq
\end{itemize}
The intersection between the curve in eq. (\ref{riccioletto}) and the end-of-the-world brane
is the point: 
\beq
(t_1, z_1, x_1)= \le \frac{l}{4} ,
 \frac{l}{4} \sin \alpha ,  - \frac{l}{4} \cos \alpha \ri \, .
\eeq 
It is useful to determine the intersection between the RT geodesic in eq.~(\ref{RT-ads-bcft})
and the brane in eq.~(\ref{posizione-brana}), which is the point with coordinates
\beq
(t_2, z_2, x_2)  =\left( 0 \, , 
 \frac{l}{2} \sin \alpha \, , 
 - \frac{l}{2} \cos \alpha \right) \, .
\label{eq:z_x_max_left_BCFT}
\eeq
The following inequalities hold
\beq
z_{1}  < z_{2}  \, , \qquad
\left| x_{1} \right| < \left| x_{2}  \right| \, .
\eeq
The intersection between the RT and the cutoff surface is the point with coordinates
\beq
(t_3,z_3,x_3)=\le 0 \, ,  \delta \, , 
 \frac{1}{2} \sqrt{l^2 - 4 \delta^2} \ri \,  .
\eeq
Note that $x_{3}$ is different from $x_{\rm int,R}$ evaluated at $z=\delta,$ which is instead
\beq
x_{\rm int,R} (\delta)= \frac{1}{2} \sqrt{l(l-4 \delta)} \, , 
\eeq
and in particular 
\beq
x_{3}  \geq x_{\rm int,R} (\delta) \, .
\eeq
Therefore, when splitting the evaluation of the gravitational action using regularization A
 in Fig.~\ref{fig-2regs}, we need to include the contribution from spacetime regions in this interval.
The full geometric setting is depicted in Fig.~\ref{fig-geometrical_setting_BCFT}.
A projection on the $(x,z)$ plane is shown in figure \ref{fig-dominio-bcft}.

\subsection{Computation of the action}
\label{sect-computation_action_BCFT-action-terms}

The evaluation of the subregion action is composed by two parts:
 the right side of the conformal diagram in Fig.~\ref{fig-geometrical_setting_BCFT},
  which is the same as in empty AdS space, and the left side where the end-of-the-world brane modifies the geometry to consider.
The former contribution for symmetry reasons is half of the subregion action evaluated in AdS$_3$ spacetime, 
which was studied  in \cite{Carmi:2016wjl,Caceres:2019pgf,Auzzi:2019vyh}:
\beq
  I_{\rm R} = \frac{L}{8 \pi G} \log \left| \frac{\tilde{L}}{L} \right| \, \frac{l}{\delta}
   + \frac{L}{4 \pi G} \log \left| \frac{2 \tilde{L}}{L} \right| \log \le \frac{\delta}{l} \ri
    - \frac{L}{4 \pi G} \log \left| \frac{\tilde{L}}{L} \right| + \frac{L \pi}{32 G} \, .
    \label{azione-mezza-ads}
\eeq
Now we proceed with the computation of the left side.

\subsubsection{Bulk term}

\begin{figure}
\centering
\includegraphics[scale=0.8]{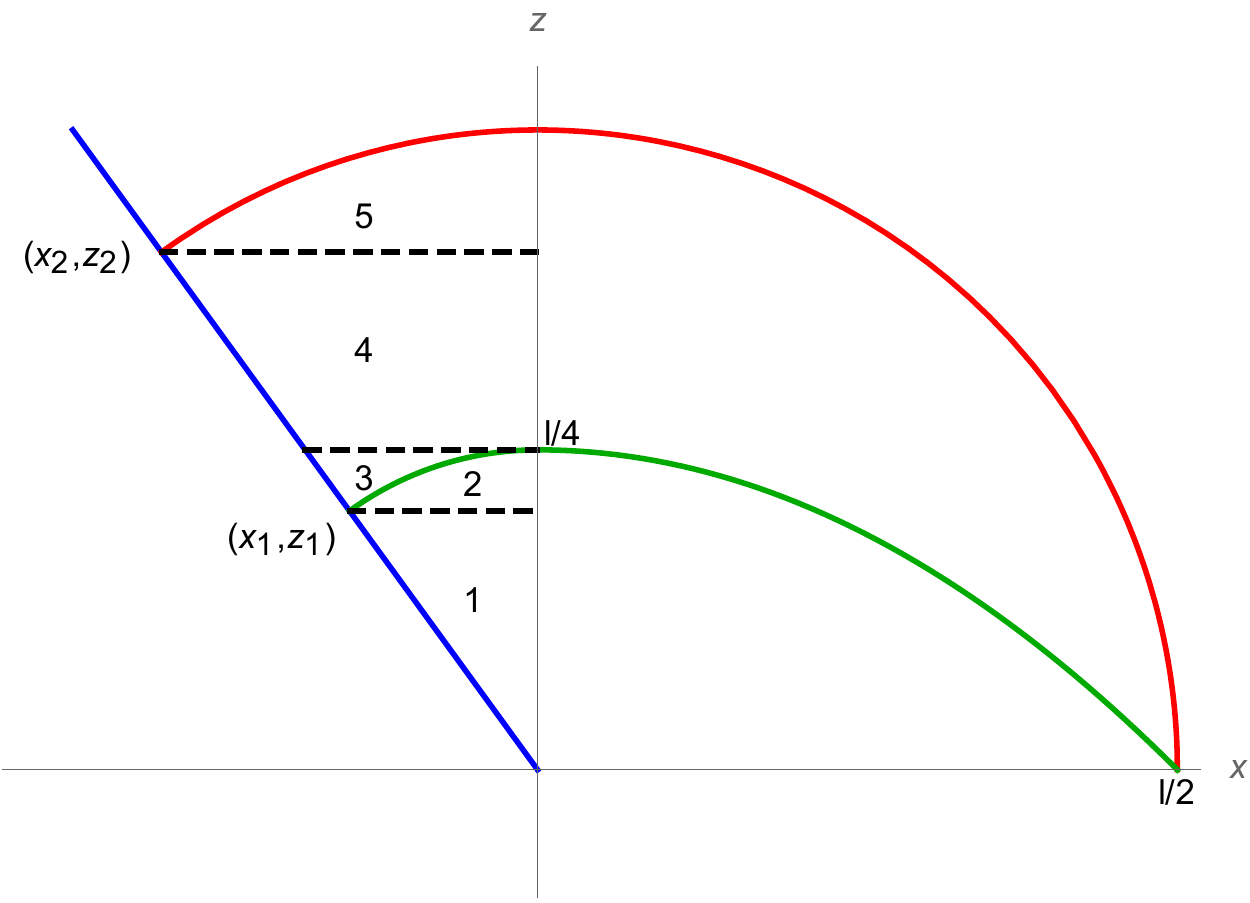}
\caption{
Projection on the $(x,z)$ plane of the integration domain.
It corresponds to figure  \ref{fig-geometrical_setting_BCFT} seen from above
(where for "above" we mean large positive $t$).
The end-of-the-world brane is shown in blue and the RT surface is drawn in red. 
The intersection curve in eqs. (\ref{riccioletto-2}) and (\ref{riccioletto}) is shown in green.
The splitting of integration in eq. (\ref{bulk-integrals-BCFT})
 for the bulk term for $\alpha \geq {\pi}/{6}$ is also shown.
 The integrals $I^{1}_{\mathcal{B},L}, \dots , I^{5}_{\mathcal{B},L} $ correspond
 to the regions $1,\dots,5$.}
\label{fig-dominio-bcft}
\end{figure}

Using the decomposition in figure \ref{fig-dominio-bcft}, for  $\alpha \geq {\pi}/{6}$ we can split
 the computation of the left side of the  bulk term as follows
\beq
I_{\mathcal{B},L} = 2 \le  I^{1}_{\mathcal{B},L} + I^{2}_{\mathcal{B},L} + I^{3}_{\mathcal{B},L} + I^{4}_{\mathcal{B},L} +  I^{5}_{\mathcal{B},L} \ri \, ,
\label{bulk-integrals-BCFT}
\eeq
where
\bea
I^1_{\mathcal{B},L} & =& - \frac{L}{4 \pi G} \int_{\delta}^{z_{1} } dz \int_{-z \cot \alpha}^{0} dx \int_{0}^{t_{\rm WDW,L}(z,x)} dt \, \frac{1}{z^{3}}  \, ,
\nl
I^2_{\mathcal{B},L} & = &- \frac{L}{4 \pi G} \int_{z_{1} }^{l/4} dz \int_{x_{\rm int,L} (z)}^{0} dx \int_{0}^{t_{\rm WDW,L}(z,x)} dt \, \frac{1}{z^{3}}  \, ,
\nl
I^3_{\mathcal{B},L} & = &- \frac{L}{4 \pi G} \int_{z_{1} }^{l/4} dz \int_{-z \cot \alpha}^{x_{\rm int,L} (z)} dx \int_{0}^{t_{\rm EW}(z,x)} dt \, \frac{1}{z^{3}}  \, , 
\label{eq:splitting_bulk_BCFT} \\
I^4_{\mathcal{B},L} & = &- \frac{L}{4 \pi G} \int_{l/4}^{z_{2}} dz \int_{-z \cot \alpha}^{0} dx \int_{0}^{t_{\rm EW}(z,x)} dt \, \frac{1}{z^{3}} \, ,
\nonumber \\
I^5_{\mathcal{B},L} & = &
- \frac{L}{4 \pi G} \int_{z_{2}}^{l/2} dz
 \int_{\sqrt{(l/2)^2-z^2}}^{0} dx
  \int_{0}^{t_{\rm EW}(z,x)} dt \, \frac{1}{z^{3}} \, .
  \nonumber
\eea
The integrals $I^{1}_{\mathcal{B},L}, \dots , I^{5}_{\mathcal{B},L} $ are performed respectively in the domains
$1,\dots,5$ shown in figure \ref{fig-dominio-bcft}.
In eq.~(\ref{bulk-integrals-BCFT}), we put a factor of 2 to account for the part of the geometry at negative times.
In the evaluation of the gravitational action, we need to consider the spacetime region given by the intersection between the EW and the WDW patch, which are delimited by their respective null boundaries.
Using the symmetry along the time direction, the recipe is to integrate along $t$ from 0 up to the smaller value between $t_{\rm WDW,L}(z,x)$ and $t_{\rm EW} (z,x).$
In the following, we will refer to the application of this prescription by stating that either the WDW patch sits below the EW, or viceversa.
Here by "below" we mean being closer to the plane $t=0$, see figure \ref{fig-geometrical_setting_BCFT}.
The intersection curves depicted in green in Fig.~\ref{fig-geometrical_setting_BCFT} and \ref{fig-dominio-bcft} delimit the region where the time coordinate of the EW patch becomes bigger than the WDW patch, or similar transitions in the splitting of the geometrical decomposition.

The first two terms in eq. (\ref{bulk-integrals-BCFT})
 refer to the region where the WDW patch is below the EW, 
and therefore the integration along $t$ goes from 0 to the function $t_{\rm WDW,L} (z,x).$
The terms $I^3_{\mathcal{B},L}, I^4_{\mathcal{B},L}, I^5_{\mathcal{B},L}$ correspond instead 
to the spacetime region where the EW sits below the WDW patch.

Strictly speaking, the decomposition in eq.~\eqref{eq:splitting_bulk_BCFT} only applies when $\alpha \geq {\pi}/{6}.$
For  $\alpha < {\pi}/{6},$ the splitting is slightly different and the integrals to evaluate get modified.
However there is nothing singular at the special value $\alpha= {\pi}/{6}:$ rather, the distinction arises from
the splitting of the integration domains. We verified by direct computation that the result for  $\alpha < {\pi}/{6}$ is 
given by the same analytical formula. 

We find that the relevant divergences all come from the first contribution
\beq
\begin{aligned}
I^1_{\mathcal{B},L} & = \frac{L}{8 \pi G} \left[ \frac{\cot \alpha}{\sin \alpha} - \log \le \tan \le \frac{\alpha}{2} \ri \ri  \right]
\left[ \log \le \frac{\delta}{l} \ri  - \log \le \frac{\sin \alpha}{4} \ri \right] \, ,
\end{aligned}
\eeq
while the other terms contribute only to finite parts.
Summing all these contributions, we get
\beq
\begin{aligned}
I_{\mathcal{B},L}  = & \frac{L}{4 \pi G} \left[ \frac{\cot \alpha}{\sin \alpha} - \log \le \tan \le \frac{\alpha}{2} \ri \ri  \right]
\left[ \log \le \frac{\delta}{l} \ri  - \log \le \frac{\sin \alpha}{4} \ri \right] \\
& + \frac{L}{96 \pi G}
\left[ \pi^2  - 36 \, \frac{\cot \alpha}{\sin \alpha} 
- 6 \log^2 2 - 12 \log^2 \le \sin \alpha \ri 
+ 6 \log \le 4-4 \cos \alpha \ri \log \le 1- \cos \alpha \ri \right. \\
& \left. 
+ 12 \log \le \tan \le \frac{\alpha}{2} \ri \ri 
- 24 \log \le 2 \sin \alpha \ri  \log \le \tan \le \frac{\alpha}{2} \ri \ri 
+ 24 \, \frac{\cot \alpha}{\sin \alpha} \, \log 2
- 12 \mathrm{Li}_2 \le \sin^2 \le \frac{\alpha}{2} \ri \ri
\right] \, .
\end{aligned}
\eeq
The divergent part is all contained into the logarithm in the first line; everything else amounts to a finite part.

\subsubsection{GHY term}

The contribution to the GHY term in the left region is evaluated exactly in the same way as in \cite{Braccia:2019xxi}, since the presence of the entanglement wedge does not modify this part.
Indeed, for small enough $\delta,$ the cutoff only intersects the WDW patch; therefore we obtain (including a symmetry factor of 2):
\beq
I_{\rm GHY,L} = \frac{L}{2 \pi G } \int_{- \delta \cot \alpha}^0 dx \int_0^{t_{\rm WDW,L} (\delta,x)} \frac{dt}{\delta^2} =
\frac{L}{4 \pi G} \left[ \frac{\cot \alpha}{\sin \alpha} - \log \le \tan \le \frac{\alpha}{2} \ri \ri  \right] \, .
\eeq
Unlike the computation in \cite{Braccia:2019xxi}, here there is no IR cutoff because the finite length of the subregion on the boundary works as a regulator.

\subsubsection{Brane term}

The brane term is a codimension-one contribution coming from the surface $\mathcal{Q}$ parametrized by $x=-z \cot \alpha.$
For the brane, the outward-directed normal vector, induced metric and extrinsic curvature are
\beq
n^{\mu} = - \frac{z}{L} \le 0, \cos \alpha, \sin \alpha \ri \, , \quad
ds^2 = \frac{L^2}{z^2} \le -dt^2 + \frac{dz^2}{\sin^2 \alpha} \ri \, , \quad
\sqrt{-h} = \frac{L^2}{z^2} \frac{1}{\sin \alpha} \, , \quad
K=  \frac{2 \cos \alpha}{L} \, ,
\label{eq:data_brane_BCFT}
\eeq
while the tension is given in eq.~\eqref{eq:tension_brane}.
The brane term can be splitted into a contribution coming from the intersection with the WDW patch, and the other one coming from the intersection with the EW.
After including a symmetry factor of 2 to account for negative times, we compute these terms as follows:
\beq
I_{\mathcal{Q}} = 2 \le I^1_{\mathcal{Q}} + I^2_{\mathcal{Q}} \ri \, , 
\eeq
where
\bea
I^1_{\mathcal{Q}} &=& \frac{1}{8 \pi G} \int_{\delta}^{z_{\rm int,L} (\alpha)} dz 
\int_0^{t_{\rm WDW,L} (z,x=-z \cot \alpha)} dt \, \frac{L}{z^2} \cot \alpha = 
\frac{L}{8 \pi G} \frac{\cot \alpha}{\sin \alpha} \log \le \frac{l}{4 \delta} \sin \alpha \ri \, ,
\nl
I^2_{\mathcal{Q}} &=& \frac{1}{8 \pi G} \int_{z_{\rm int,L} (\alpha)}^{z_{\rm max,L} (\alpha)} dz 
\int_0^{t_{\rm EW} (z,x=-z \cot \alpha)} dt \, \frac{L}{z^2} \cot \alpha = 
 \frac{L}{8 \pi G} \frac{\cot \alpha}{\sin \alpha} \le 1- \log 2 \ri \, .
\eea
Summing the two contributions, we find
\beq
I_{\mathcal{Q}} = \frac{L}{4 \pi G} \frac{\cot \alpha}{\sin \alpha} \left[ \log \le \frac{l}{\delta} \ri
+ 1 + \log \le \frac{\sin \alpha}{8} \ri
\right] \, .
\label{eq:brane_term_AdS_BCFT}
\eeq

\subsubsection{Joint terms}

There are several joint terms to include:
\begin{itemize}
\item \textbf{Joint between the cutoff surface $z=\delta$ and the brane $\mathcal{Q}.$} \\
This joint involves two timelike surfaces. The induced metric is determined by imposing $z=\delta$ and $x=-z \cot \alpha,$ so that
\beq
ds^2 = - \frac{L^2}{z^2} dt^2 \, , \qquad
\sqrt{-h} = \frac{L}{\delta} \, .
\eeq
In addition, the argument of the integrand corresponds to the boost parameter relating the cutoff and the brane, and reads
\beq
\eta = \left| \mathrm{arccos} \, \le \mathbf{n} \cdot \mathbf{n}_{\delta} \ri \right| = \alpha \, ,
\eeq
where $\mathbf{n}$ is defined in eq.~\eqref{eq:data_brane_BCFT} and $\mathbf{n}_{\delta}$ is the normal one-form to the cutoff surface.
This joint is fully contained in the region where the WDW patch sits below the EW, and therefore the integration along the coordinate $t$ runs along the range $[0, t_{\rm WDW,L} (z=\delta, x=-\delta \cot \alpha)].$
Explicitly, this is given by
\beq
I_{\mathcal{J}, \rm L}^{\mathcal{Q},\delta} = \frac{1}{4 \pi G} \int_0^{\frac{\delta}{\sin \alpha}} dt \, \frac{\alpha L}{\delta} = 
\frac{L}{4 \pi G} \frac{\alpha}{\sin \alpha} \, .
\eeq
We included here a symmetry factor of 2 to include also the region at negative times.
This expression is the same appearing in \cite{Braccia:2019xxi}.
\item \textbf{Joint between the WDW patch and the cutoff surface $z=\delta.$} \\
The WDW patch at $z=\delta$ is given by $t_{\rm WDW,L} (\delta, x) = \sqrt{x^2+\delta^2}.$ 
Therefore the induced metric reads
\beq
ds^2 = \frac{L^2}{z^2} \left[ - \le \frac{dt_{\rm WDW,L}}{dx} \ri^2 +1  \right] dx^2 \, , \qquad
\sqrt{\gamma} = \frac{L}{\cos \theta} \, ,
\eeq
where we are using for convenience polar coordinates such that $\theta = - \mathrm{\arctan} \, (x/z).$
The scalar product between the normal one-forms is
\beq
\mathfrak{a} = \log \left| \mathbf{n}_{\delta} \cdot \mathbf{k}_{\rm WDW,L} \right| = \log \left| \frac{B L \cos \theta}{\delta}  \right| \, .
\eeq
This allows to evaluate
\beq
\begin{aligned}
I_{\mathcal{J}, \rm L}^{\rm WDW, \delta} & = 
 - \frac{L}{4 \pi G}  \int_0^{\frac{\pi}{2}-\alpha} \frac{d\theta}{\cos \theta} \log \left| \frac{B L \cos \theta}{\delta}  \right| = \\
 & = \frac{L}{4 \pi G} \log \le \tan \le \frac{\alpha}{2} \ri \ri \log \le \frac{B L}{\delta} \ri \\
 &  - \frac{L}{96 \pi G} \left[ \pi^2 + 12 \log \le \tan \le \frac{\alpha}{2} \ri \ri  \log \le \frac{1}{4} \tan \le \frac{\alpha}{2} \ri \ri 
 + 12 \mathrm{Li}_2 \le - \cot^2 \le \frac{\alpha}{2} \ri \ri  \right]   \, .
\end{aligned}
\eeq
where we assigned $\eta_{\eta}=-1$ sign due to the fact that the joint is a past boundary for the WDW patch, and the spacetime region of interest is also in the past with respect to the null boundary of the WDW patch.
In addition, we put a symmetry factor of 2.
There would be in principle a joint term from the intersection between the cutoff surface at $z=\delta$ and the EW, 
but since the brane delimits the spacetime region of integration, this intersection is cut away.
\item \textbf{Joint at the RT surface.} \\
The computation works in the same way as for empty AdS spacetime, the only difference being that the lower endpoint of integration along $x$ is determined by putting $x= - z \cot \alpha$ in the equation defining the RT surface.
We get
\bea
I_{\mathcal{J}, \rm L}^{\rm RT} & =& - \frac{L}{8 \pi G} \int_{-\frac{l}{2} \cos \alpha}^0 dx
 \, \frac{2l}{l^2-4x^2} \log \left| \frac{C^2}{4 L^2} \le l^2 - 4 x^2 \ri  \right|  \nl
& = & \frac{L}{4 \pi G} \log \le \tan\le \frac{\alpha}{2} \ri \ri
\log \le \frac{C l}{L} \ri 
\nl &&
- \frac{L}{96 \pi G} 
\left[ \pi^2 + 12 \log^2 \le \tan \le \frac{\alpha}{2} \ri \ri
+ 12 \mathrm{Li}_2 \le -\cot^2 \le \frac{\alpha}{2} \ri \ri  \right]
 \, .
\eea
We notice that this joint amounts only to a finite part.
\item \textbf{Joint at the intersection curve between WDW patch and EW.} \\
The induced metric is
\beq
ds^2 = \frac{L^2}{z^2} \left[  \le \frac{dz_{\rm int,L}}{dx} \ri^2 +1  \right] dz^2 = 
\frac{L^2}{z_{\rm int,L}^2}\frac{l^2}{l^2-16x^2} dz^2 \, , \qquad
\sqrt{\gamma} =  \frac{4l}{l^2-16x^2} \, ,
\eeq
and the integrand contains the factor
\beq
\mathfrak{a} = \log \left| \mathbf{k}_{\rm WDW,L} \cdot \mathbf{w}  \right| = 
\log \left| B C   \right| \, .
\eeq
In this computation, we used the fact that the intersection curve is located at constant $t=l/4.$
Therefore we obtain (putting a symmetry factor of 2)
\beq
\begin{aligned}
I_{\mathcal{J}, \rm R}^{\rm int}  = \frac{L}{4 \pi G} \int_{-\frac{l}{4} \cos \alpha}^{0} dx \,  \frac{4l}{l^2-16x^2} \log \le B C \ri = 
 - \frac{L}{4 \pi G} \log \le B C \ri  \log \le \tan \le \frac{\alpha}{2} \ri \ri \, . 
\end{aligned}
\eeq
\item \textbf{Joint between orthogonal surfaces} \\
There are three joints between orthogonal surfaces. One of them is between the
  two parts of the WDW patch at $x=0$ and involves two parallel 
 null surfaces. The other two involve the joints between the brane and the WDW patch and the EW, respectively.
All of them have a potentially divergent integrand,
which involve the logarithm of the scalar product of two orthogonal vectors.
However, the integrand is multiplied by the  vanishing determinant of the induced metric.
Although formally undetermined, these terms can be shown to vanish
by an opportune limiting procedure  \cite{Braccia:2019xxi}.
\end{itemize}
Summing all the joint contributions from the left side of the geometry, we obtain
\beq
\begin{aligned}
I_{\mathcal{J}, \rm L} & = 
- \frac{L}{4 \pi G} \log \le \tan \le \frac{\alpha}{2} \ri \ri 
 \log \le \frac{\delta}{l} \ri \\
& + \frac{L}{48 \pi G} 
\left[ 12 \, \frac{\alpha}{\sin \alpha} - 12 \log \le \tan \le \frac{\alpha}{2} \ri \ri \log \le \frac{1}{2} \tan \le \frac{\alpha}{2} \ri \ri 
- 12 \mathrm{Li}_2 \le - \cot^2 \le \frac{\alpha}{2} \ri \ri
-  \pi^2 \right] \, .
\end{aligned}
\eeq
The only divergence is logarithmic and comes from the joint between the WDW patch and the cutoff surface; all the other terms contribute to finite parts.
Remarkably, all the dependence on the ambiguity in parametrizing the null vectors cancel, and we obtain a logarithmic divergence in $\delta/l$ plus a finite part which depends only on $\alpha.$

\subsubsection{Null boundary term and counterterm}

The computation of the codimension-one terms on the WDW patch works in the same way as explained in \cite{Sato:2019kik, Braccia:2019xxi}, except that the range of the $\lambda$ coordinate is delimited by the intersection between the WDW patch and the EW.
In particular, the biggest value of $z$ reached along the null boundary of the WDW patch is $z=l/4.$ 
We find
\beq
I_{\rm \mathcal{N}}^{\rm WDW,L} = - \frac{1}{4 \pi G} \int_0^{\frac{\pi}{2}-\alpha} d \theta
 \int_{\frac{\delta}{B \cos \theta}}^{\frac{l}{4B \cos \theta}} d \lambda \, \frac{L}{\cos \theta} 
 \le - \frac{2}{\lambda} \ri = \frac{L}{2 \pi G} \log \le \tan \le \frac{\alpha}{2} \ri  \ri
 \log \le \frac{4 \delta}{l} \ri \, .
\eeq
Since the expansion parameter of the congruence of geodesics 
vanishes both on the left side of the WDW patch and on the
boundary of the EW, the counterterm action is zero, 
$I_{\rm ct} = 0$.

One should also consider the analogous terms evaluated on the EW; however they both vanish because the parametrization \eqref{eq:null_vector_EW_AdS_BCFT} is affine and the expansion parameter vanishes.

\subsubsection{Subtraction of the vacuum solution and final result}
\label{sect-subtraction_vacuum_BCFT}

Summing all the contributions, one observes that the divergences arising  from 
the left side cancel, giving just a finite contribution.
There is also a contribution to the total action from the right side,
(\ref{azione-mezza-ads}), which is independent of $\a$.

\begin{figure}[ht]
\centering
\includegraphics[scale=1.2]{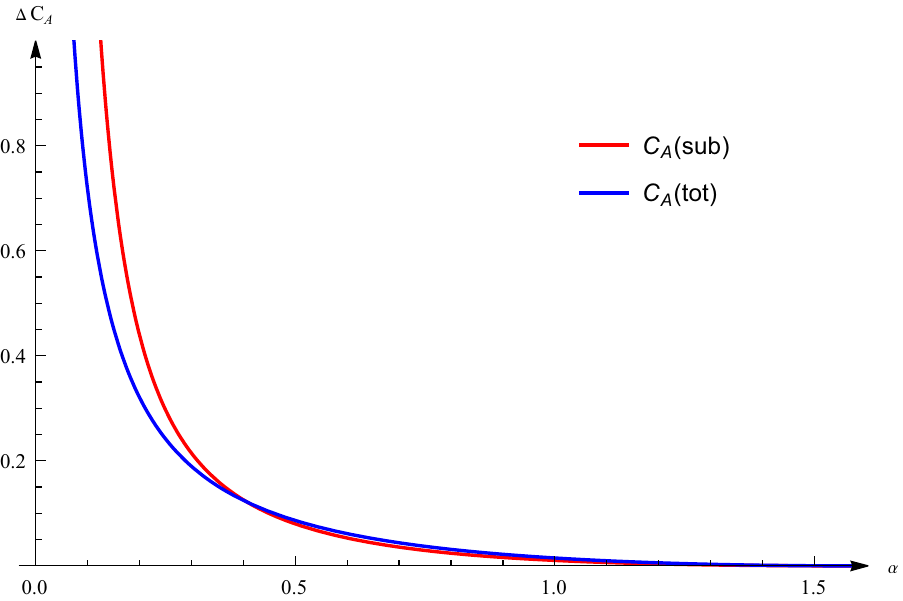}
\caption{Plot of the   contribution of the defect to the subregion complexity  in eq.~\eqref{eq:complexity_formation_BCFT_section5} 
as a function of $\alpha$ (red) and comparison with the result for the total space studied in \cite{Sato:2019kik} (blue). 
Here we set $L/G=1$.}
\label{fig-complexity_formation_BCFT}
\end{figure}

As in  \cite{Sato:2019kik}, we can isolate the contribution of the boundary
by subtracting the complexity of the $\alpha=\frac{\pi}{2}$ 
case\footnote{Note that the conventions used in \cite{Sato:2019kik} are such that their parameter, 
that we call $\tilde{\alpha},$ is related to the $\alpha$ used here and in \cite{Braccia:2019xxi} 
by the relation $\tilde{\alpha}= \cot \alpha .$},
which correspond to a vanishing brane tension, see eq. (\ref{eq:tension_brane}).
We find that the boundary contribution to the subregion complexity is finite,
and independent from $l$ i.e.
\beq
 \begin{aligned}
  \Delta \mathcal{C}_A^{\rm BCFT}  =   \frac{L}{96 \pi^2 G}  & \left\lbrace  
  12  \frac{\cot \alpha}{\sin \alpha} - \pi^2 - 6 \log^2 2 -12 \log^2 \le \sin \alpha \ri 
  + 24 \, \frac{\alpha}{\sin \alpha} -12 \log \le \tan \le \frac{\alpha}{2} \ri \ri   \right. \\
  & \left.   
  + 6 \log \le 4- 4 \cos \alpha \ri \log \le 1- \cos \alpha \ri 
  + 24 \log \le \tan \le \frac{\alpha}{2} \ri \ri \log 4
  \right. \\
&   \left.  - 24 \log^2 \le \tan \le \frac{\alpha}{2} \ri \ri  
  - 12 \mathrm{Li}_2 \le \sin^2 \le \frac{\alpha}{2} \ri \ri 
- 24 \mathrm{Li}_2 \le  - \cot^2 \le \frac{\alpha}{2} \ri \ri  \right\rbrace 
  \, .
\end{aligned} 
\label{eq:complexity_formation_BCFT_section5}
\eeq
The result is plotted in Fig.~\ref{fig-complexity_formation_BCFT},
 showing a divergent result when $\alpha \rightarrow 0$ and its vanishing value when $\alpha= \frac{\pi}{2}.$
 
As the contribution of the defect $  \Delta \mathcal{C}_A^{\rm BCFT}  $ to the subregion complexity is independent of the subregion size $l$, we may expect that it should reproduce the calculation of the total complexity
of formation in \cite{Braccia:2019xxi,Sato:2019kik}. This is not the case,
because the choice of the infrared cutoff is different: while in
\cite{Braccia:2019xxi,Sato:2019kik} the action is regulated by an IR
cutoff at constant $z$, we instead use as an IR cutoff the RT surface.
The two choices agree just for the UV divergent part of the 
action: in fact, this divergence is independent of the infrared regulator
because it is localised nearby the location  of the defect.


\section{Conclusions}
\label{sect-conclusion}

In this work we computed the CA conjecture for a subregion given by an interval of length $l$ 
 on the boundary for both the Janus AdS$_3$ geometry and the AdS$_3$/BCFT$_2$ model.
As discussed below Table \ref{tab:results}, the action conjecture does not provide 
a universal structure of UV divergences, but the results depend on the particular defect or boundary characterizing the geometry.
It was recently proposed that ambiguities in the field theory realization of 
complexity models could be related to similar ambiguities in the holographic proposal \cite{Belin:2021bga}.
It would be interesting to investigate if the distinct behaviours of the interface models considered in this paper can be related to such ambiguities. 

In \cite{ Baiguera:2021cba} we studied the volume conjecture for the non-supersymmetric
Janus AdS$_5$ geometry. We computed the volume using the single and the double cutoff regularizations,
and we found that only the coefficient of the log-divergences was independent of the regularization. 
It would be interesting to check this also for action complexity.

We believe that further insights on the universality properties of the holographic complexity conjecture
 could arise from an investigation of the subregion action in the moving mirror model,
  which was studied using the CV conjecture \cite{Sato:2021ftf}.
The moving mirror setting is also a useful tool to investigate the evaporation of black holes. 
The recent developments coming from the island conjectures \cite{Almheiri:2019hni,Almheiri:2019psf,Penington:2019npb} 
have been related to subregion complexity in  \cite{Hernandez:2020nem}.
It would be interesting to test these conjectures in the models with defects or boundaries considered here.


\section*{Acknowledgements}

We thank K. Toccacelo for collaboration in the first stages of the work and S. Chapman for useful discussions.
The authors S.B. acknowledge support from the Israel Science Foundation (grant No. 1417/21),
 the Kreitmann School of Advanced Graduate Studies, the Independent Research Fund Denmark
  grant number DFF-6108-00340 “Towards a deeper understanding of black holes 
  with non-relativistic holography” and the DFF-FNU through grant number DFF-4002-00037.


\appendix

\section{Jacobi elliptic functions and elliptic integrals}
\label{app-elliptics}

We follow the conventions of \cite{Abram1964} to define the incomplete elliptic integrals
\begin{align}
\elF{\vf}{m}&=\int_0^\vf\frac{d\th}{\sqrt{1-m\sin^2\th}}\;,  \noindent \\
\elE{\vf}{m}&=\int_0^\vf d\th\sqrt{1-m\sin^2\th}\;,\noindent \\
\Pi\left(n;\vf \left| m \right.\right) &=\int_0^\vf \frac{d\th}{\left(1-n\sin^2\th\right)\sqrt{1-m\sin^2\th}}.
\end{align}
of the first, second and third kind, respectively. The complete elliptic integrals are defined as
\beq
\elF{\frac{\pi}{2}}{m}=\eK(m)\;, \quad \elE{\frac{\pi}{2}}{m}=\eE(m)\;, \quad \Pi\left(n; \frac{\pi}{2}\Big| m \right)=\eP{n}{m}\;.
\eeq 
The Jacobi amplitude $\vf=\text{am}(x|m)$
 is the inverse of $\elF{x}{m}$
\beq
x=\elF{\vf
}{m}\;.
\eeq
The Jacobi elliptic functions are defined as
\beq
\label{Jels}
\sn{x}{m}=\sin\vf,\quad\cn{x}{m}=\cos\vf\quad\text{and}\quad\dn{x}{m}=\sqrt{1-m\sin^2\vf},
\eeq
such that $\sn{\mathds{K}(m)}{m}=1$ and $\cn{\mathds{K}(m)}{m}=0$. 
They satisfy the identities
\bea
& \sn{x}{m}^2+\cn{x}{m}^2=1\;, & \\
& \dn{x}{m}^2+ m \, \sn{x}{m}^2=1\;. &
\eea

We conclude with the derivation of the finite form of the change of variables presented in eq.~\eqref{tanh-sn-relazione}.
Infinitesimally, we apply the transformation in eq.~\eqref{eq:Janus_metric}. Taking 
eq.~\eqref{eq:solutions_f_dilaton_mu_coordinates}  into account, a direct integration leads to
\beq
\begin{aligned}
y &=  \int_0^{\mu} ds \, \sqrt{f(s)}  \\
& =\int_0^{\mu} ds \, \frac{\alpha_+}{\mathrm{sn} \le \alpha_+ (s + \mu_0) | m \ri} \\
& =\int_0^{\alpha_+ \mu} d\sigma \, \frac{1}{\mathrm{sn} \le \sigma + \mathds{K} (m) | m \ri} \\
& =\int_0^{\alpha_+ \mu} d\sigma \, \frac{\mathrm{dn } \le \sigma | m \ri}{\mathrm{cn }\le \sigma | m \ri} \\
&= \frac12 \mathrm{log}\left( \frac{1+ \mathrm{sn} \le \alpha_+ \mu | m \ri}{1-\mathrm{sn} \le \alpha_+ \mu | m \ri} \right)\\
&= \mathrm{arctanh}  \left[ \mathrm{sn} \le \alpha_+ \mu | m \ri \right] \, .
\end{aligned}
\label{derivazione-eq-3.40}
\eeq
In the first equation we replaced eq.~\eqref{eq:solutions_f_dilaton_mu_coordinates}, in the 
second we used the definition of $\mu_0$ in eq.~\eqref{eq:data_elliptic_functions}; then the (half)-periodicity property  $\mathrm{sn} \le \sigma + \mathds{K} (m) | m \ri =\mathrm{cn} \le \sigma | m \ri /\mathrm{dn} \le \sigma | m \ri$ was taken into account. The resulting (indefinite) integral is known and can be found for example in 
eq. $(5.135.5)$ of \cite{gradshteyn2014table}.
Inverting the hyperbolic  function leads to the form reported in eq.~\eqref{tanh-sn-relazione}.


\section{Details of the series expansion of Janus AdS action}
\label{app-details_expansion_Janus_AdS}

We report in this Appendix the details of the series-expansion of the terms composing the gravitational action evaluated in section \ref{sect-computation_action_Janus},
 following the steps outlined in section \ref{sect-technique_series_expansion_action_JAdS}.

Using the definition of $\mu_0$ in eq. (\ref{eq:data_elliptic_functions})
  and the property $f(\mu^*(\varepsilon)) = \varepsilon^{-2},$ we obtain (see eq. $(2.12)$ in \cite{Bak:2007jm})
\beq
\mu_0 - \mu^*(\varepsilon) =
 \frac{1}{\alpha_+}  \elF{  \mathrm{arcsin} \le \alpha_+ \varepsilon \ri}{ m} = \varepsilon  + \mathcal{O} (\varepsilon^3)
  \, .
 \label{eq:expansion_mu_star}
\eeq
This identity is the building block to determine the expansion of the action around $\varepsilon=0,$ 
and it is sufficient to perform the expansion of the terms of case 1 in Section \ref{sect-technique_series_expansion_action_JAdS}.

In order to perform the procedure described for the terms of case 2  in Section \ref{sect-technique_series_expansion_action_JAdS},
 the following Laurent-expansions around $\mu=\mu_0$ are needed:
\beq
f(\mu)^{3/2} \, \sin \le \mu_0 - \mu \ri = 
\frac{1}{\le \mu_0 - \mu \ri^2} + \frac{1}{3} + \mathcal{O} (\mu_0 - \mu) \, ,
\label{eq:first_Laurent_expansion}
\eeq
\beq
f(\mu)^{3/2} \, \left[ \sin \le \mu_0 - \mu \ri +1 \right] \, \log \left[ \sin \le \mu_0 - \mu \ri +1    \right] = 
\frac{1}{\le \mu_0-\mu \ri^2} + \frac{1}{2 \le \mu_0 - \mu \ri} + \frac{1}{6} + \mathcal{O} \le \mu_0 - \mu \ri \, ,
\eeq
\beq
\sqrt{f(\mu)} \log \left[  \sqrt{f(\mu)} \le  \sin \le \mu_0 - \mu \ri +1 \ri  \right]  =
- \frac{1}{\mu_0 - \mu}  \log  \le  \mu_0 - \mu \ri  +1 + \mathcal{O} \left[ \le \mu_0 - \mu \ri \log \le \mu_0 - \mu \ri \right] \, .
\eeq
In the following subsection we perform the procedure 
 term by term.

\subsection{Expansion of the action term by term}
\label{sect-app-expansion_term_by_term}
Using the identities and the definitions listed above, we determine the expansions of the terms entering the gravitational action:
\begin{itemize}
\item \textbf{Bulk term.} 
Consider the second term in eq.~\eqref{eq:bulk_term_JAdS_before_exp}, which is analytic in all the integration domain, except for a neighbourhood of $\mu=\mu_0.$
Its Laurent-expansion around $\mu=\mu_0$ is given in eq.~\eqref{eq:first_Laurent_expansion}.
The only term contributing to a divergence is the first one, therefore we regularize the integral by adding and subtracting such term, which evaluates to
\beq
\int_0^{\mu^*(\varepsilon)} \frac{d\mu}{\le \mu_0-\mu \ri^2} = \frac{1}{\mu_0 - \mu^*(\varepsilon)} - \frac{1}{\mu_0}= \frac{1}{\varepsilon} - \frac{1}{\mu_0} + \mathcal{O} (\varepsilon) \, 
\label{eq:first_divergence_I_I1}
\eeq
where we used the definition of $\mu_0$ in eq.~\eqref{eq:data_elliptic_functions} and the property $f(\mu^*(\varepsilon)) = \varepsilon^{-2}$ to obtain the expansion around $\varepsilon=0.$
Now we move to the first part of the integral \eqref{eq:bulk_term_JAdS_before_exp}.
The series expansion of the integrand around $\mu=\mu_0,$ which is the only point where singularities arise, reads
\beq
f(\mu)^{3/2} \, \left[ \sin \le \mu_0 - \mu \ri +1 \right] \, \log \left[ \sin \le \mu_0 - \mu \ri +1    \right] = 
\frac{1}{\le \mu_0-\mu \ri^2} + \frac{1}{2 \le \mu_0 - \mu \ri} + \frac{1}{6} + \mathcal{O} \le \mu_0 - \mu \ri \, .
\eeq
In this case there are two divergent terms: the first one corresponds precisely to Eq.~\eqref{eq:first_divergence_I_I1}, while the second one is computed as follows:
\beq
\int_0^{\mu^*(\varepsilon)} \frac{d\mu}{2 \le \mu_0 - \mu \ri} = \frac{1}{2} \log \le \frac{\mu_0}{\mu_0 - \mu^*(\varepsilon)} \ri 
= - \frac{1}{2} \log \le \frac{\varepsilon}{\mu_0}  \ri + \mathcal{O} (\varepsilon^2) \, .
\eeq
We can now combine all the previous results to find
\beq
I_{\mathcal{B}} (\gamma) = \frac{L}{\pi G} \left\lbrace \log \le \frac{2 \delta}{l} \ri  \le \frac{1}{\varepsilon} - \frac{1}{\mu_0} \ri 
- \frac{1}{2} \log \le \frac{\varepsilon}{\mu_0} \ri 
+ \mathcal{I}_{\mathcal{B}}^{(0)} (\gamma) + \left[ \log \le \frac{2 \delta}{l} \ri -1 \right] \mathcal{I}_{\mathcal{B}}^{(1)} (\gamma)
\right\rbrace + \mathcal{O} (\varepsilon) \, ,
\label{eq:bulk_term_expanded}
\eeq
where $\mathcal{I}_{\mathcal{B}}^{(0)}$ and $\mathcal{I}_{\mathcal{B}}^{(1)}$
are given in
 \eqref{eq:numerical_function_IB0} and \eqref{eq:numerical_function_IB1}.
\item \textbf{GHY term.}
The GHY term arising from the cutoff surface located at $z=\delta$ and evaluated in eq.~\eqref{eq:GHY_term1_JAdS_before_exp} is already a finite expression, which we denote as the numerical function \eqref{eq:numerical_function_IGHY}.
The GHY contribution coming from the surfaces at $\mu=\pm \mu^*(\varepsilon)$ are expanded by means of the identity \eqref{eq:expansion_mu_star}.
The sum of both terms is given by
\beq
I_{\rm GHY} (\gamma) =  -\frac{L}{\pi G} \left[ \frac{1}{\e}\log\left(\frac{2\d}{l} \right) +\frac{1}{2}
 - \frac{1}{2} \, \mathcal{I}_{\rm GHY} (\gamma)  \right] + \mathcal{O} (\varepsilon) \, ,
\label{eq:GHY_expanded}
\eeq
where $\mathcal{I}_{\rm GHY} (\gamma)$ is given in eq. (\ref{eq:numerical_function_IGHY}).
\item \textbf{Null boundary term.}
We consider the expression \eqref{eq:null_boundary_term_JAdS_before_exp} and perform an integration by parts of the second term (which we report here for convenience) to get
\beq
\begin{aligned}
& \frac{L}{\pi G} \int_0^{\mu^* (\varepsilon)} d\mu \, \frac{f'(\mu)}{2 \sqrt{f(\mu)}} \cos \le \mu_0 - \mu \ri
 \le
\log \le \frac{l}{2 \delta} \ri - \log  \le 1+ \sin \le \mu_0 - \mu \ri \ri  \ri   =  \\
& = \frac{L}{\pi G} \left[ \sqrt{f(\mu)} \cos \le \mu_0 - \mu \ri \le
\log \le \frac{l}{2 \delta} \ri - \log  \le 1+ \sin \le \mu_0 - \mu \ri \ri  \ri  \right]_0^{\mu^*(\varepsilon)}  \\
& - \frac{L}{\pi G} \int_0^{\mu^*(\varepsilon)} d\mu \, \sqrt{f(\mu)} 
\left\lbrace 1 + \sin \le \mu_0 - \mu \ri \le \log \le \frac{l}{2 \delta} \ri - \log  \le 1+ \sin \le \mu_0 - \mu \ri \ri  -1  \ri \right\rbrace \, .
\end{aligned}
\label{eq:preliminar_expansion_null_boundary_term}
\eeq
The series expansion of the part without any further integration is simplified by the properties $f(0) = \alpha_+^2$ and $f(\mu^*(\varepsilon))=\varepsilon^{-2}.$
Instead the last line combines with the first term in eq.~\eqref{eq:null_boundary_term_JAdS_before_exp} to a simpler expression:
\beq
\frac{L}{\pi G} \int_0^{\mu^*(\varepsilon)} d\mu \, \sqrt{f(\mu)} \left[ \sin \le \mu_0 - \mu \ri -1 \right] = 
 \frac{L}{\pi G} \, \mathcal{I}_{\rm GHY} (\gamma) -  \frac{L}{\pi G} \int_0^{\mu^*(\varepsilon)} d\mu \, \sqrt{f(\mu)} \, .
\eeq
The last contribution is  evaluated using the change of variables $\sqrt{f(\mu)} d\mu = dy.$
Collecting all the results, we get
\beq
\begin{aligned}
I_{\mathcal{N}} (\gamma)  = 
 \frac{L}{\pi G} & \left\lbrace  \frac{1}{\varepsilon} \log \le \frac{l}{2 \delta} \ri - \alpha_+ \cos \mu_0 \log \le \frac{l}{2 \delta} \ri
 + \alpha_+ \cos \mu_0 \log \le 1+\sin \mu_0 \ri  \right. \\
& \left. +  \log \le (1-2\gamma^2)^{1/4} \,  \frac{\varepsilon}{2} \ri 
  + \mathcal{I}_{\rm GHY} (\gamma) 
    -1  \right\rbrace + \mathcal{O} (\varepsilon) \, .
    \label{eq:I_N_expanded}
\end{aligned}
\eeq 
\item \textbf{Joint terms.}
Most of the joint terms can be evaluated immediately using the identity \eqref{eq:expansion_mu_star} and the numerical functions defined in Appendix \ref{app-collection_num_functions}.
We also need this explicit integral, giving a divergent part:
\beq
\begin{aligned}
&  \int_0^{\mu^*(\varepsilon)}  \frac{d \mu}{\mu_0 - \mu} \log \le \mu_0 - \mu \ri    =
 - \frac{1}{2} \le \log^2 \varepsilon - \log^2 \mu_0  \ri  
 + \mathcal{O} (\varepsilon) \, .
 \end{aligned}
\eeq
The result for the total joint action is
\beq
I_{\mathcal{J}} (\gamma)= I^{\delta, \rm WDW}_{\mathcal{J}} (\gamma) 
+ I^{\varepsilon, \rm WDW}_{\mathcal{J}} (\gamma) +
I_{\mathcal{J}}^{\rm RT}  (\gamma) + I_{\mathcal{J}}^{\rm int} (\gamma) +
I^{\mu}_{\mathcal{J}} (\gamma) \, ,
\label{eq:IJ_expanded}
\eeq
where
\beq
I^{\delta, \rm WDW}_{\mathcal{J}} (\gamma) = 
- \frac{L}{2\pi G}\left[ \log\left( \frac{\alpha L}{\delta}\right) \mathcal{I}_{\rm GHY} (\gamma)  + \mathcal{I}_{\mathcal{J}}^{(0)} (\gamma) \right]\, ,
\label{eq:IJ1_expanded}
\eeq
\beq
\hspace{-\leftmargin}
I^{\varepsilon, \rm WDW}_{\mathcal{J}} (\gamma) = \frac{L}{4 \pi G}  \left\lbrace
\frac{1}{\varepsilon} \left[ 2 \log \le \frac{\alpha L}{\varepsilon}  \ri  \log \le  \frac{2 \delta}{l} \ri
- \log^2 \delta  + \log^2 \le \frac{2}{l} \ri \right]
+ 2 \log \le \frac{2 \alpha L}{l \varepsilon}  \ri
\right\rbrace  + \mathcal{O} (\varepsilon)  \, ,
\label{eq:IJ2_expanded}
\eeq
\beq
\hspace{-\leftmargin}
I_{\mathcal{J}}^{\rm RT}  (\gamma) + I_{\mathcal{J}}^{\rm int} (\gamma) =  \frac{L}{2 \pi G} \left\lbrace   
\frac{1}{2} \log^2 \varepsilon 
-  \log \le  \frac{\alpha L}{l} \ri \log \le \le 1- 2 \gamma^2 \ri^{1/4} \frac{\varepsilon}{2} \ri 
- \frac{1}{2} \log^2 \mu_0 
 +  \, \mathcal{I}_{\mathcal{J}}^{(1)} (\gamma) \right\rbrace
 + \mathcal{O} (\varepsilon) \, ,
 \label{eq:IJ3_expanded}
\eeq
\beq
\begin{aligned}
I^{\mu}_{\mathcal{J}} (\gamma) & =
 - \frac{L \alpha_+ \, \cos \mu_0}{4 \pi G}
\left\lbrace  
2 \log | \alpha L \alpha_+ \cos \mu_0 |
\log \le \frac{2 \delta}{l} \ri  \right. \\
& \left.
 + 2 \log \left| \alpha L \alpha_+ \cos \mu_0  \right|
\log \le 1+ \sin \mu_0 \ri
- \log^2 \delta 
+  \log^2 \le \frac{2(1+ \sin \mu_0)}{l} \ri
 \right\rbrace  + \mathcal{O} (\varepsilon)  \, ,
\end{aligned}
\label{eq:IJ4_expanded}
\eeq
where the functions $ \mathcal{I}_{\mathcal{J}}^{(0)} (\gamma) $ 
and $ \mathcal{I}_{\mathcal{J}}^{(1)} (\gamma) $ are respectively defined in
eqs. (\ref{eq:numerical_function_J0}) and (\ref{eq:numerical_function_J1}).
It is relevant to observe that the ambiguity in the normalization of the null normals to the boundary of the EW, 
parametrized by $\beta,$ cancels once we combine the joints at the RT surface and at the intersection between WDW patch and EW.
\item \textbf{Counterterm.}
The treatment of the counterterm is similar to the null boundary term; since the expression is rather cumbersome, we directly report the result:
\beq
\hspace{-\leftmargin}
\begin{aligned}
 I_{\rm ct}^{\rm WDW} (\gamma) & = 
\frac{L}{4 \pi G} \left\lbrace   
  2 \log \le \frac{2 \alpha \tilde{L}}{l} \ri \log \le (1-2 \gamma^2)^{1/4} \, \frac{\varepsilon}{2} \ri 
+ 2 \log \le \frac{2 \alpha \tilde{L}}{l} \ri \mathcal{I}_{\rm GHY} (\gamma) 
+ \log^2 \mu_0 - \log^2 \varepsilon \right.  \\
& \left.  + \mathcal{I}_{\rm ct}^{(0)} (\gamma) 
+ 2 \log \le \frac{2 \delta}{l} \ri  \le \frac{1}{\varepsilon} - \frac{1}{\mu_0} \ri  +  \log \le \frac{2\delta}{l} \ri \mathcal{I}_{\rm ct}^{(1)} (\gamma)  
+ 2 \log \le \frac{\mu_0}{\varepsilon} \ri   +  \mathcal{I}_{\rm ct}^{(2)} (\gamma)  \right. \\
& \left.
- \frac{1}{\varepsilon} \left[
2 \log \le \frac{\alpha \tilde{L}}{\varepsilon} \ri
\log \le \frac{2 \delta}{l} \ri
- \log^2 \delta 
+ \log^2 \le \frac{2}{l} \ri \right] 
- 2 \log \le \frac{2 \alpha \tilde{L}}{l \varepsilon} \ri  \right. \\
& \left. + \alpha_+ \cos \mu_0 
\left[ 2 \log \le \alpha \tilde{L} \sin \mu_0 \ri \log \le \frac{2 \delta}{l} \ri
+ 2 \log \le \alpha \tilde{L} \sin \mu_0 \ri  \log \le 1+\sin \mu_0 \ri \right] \right. \\
& \left. + \alpha_+ \cos \mu_0 \left[  \log^2 \le \frac{2 \le 1+\sin \mu_0 \ri}{l} \ri - \log^2 \delta  \right]
\right\rbrace + \mathcal{O}(\varepsilon)  \, ,
\end{aligned}
\label{eq:Ict_expanded}
\eeq
where the numerical functions $ \mathcal{I}_{\rm ct}^{(0)}, \mathcal{I}_{\rm ct}^{(1)} $ ar defined in eqs.~\eqref{eq:numerical_function_Ict0}  and \eqref{eq:numerical_function_Ict1}.

\end{itemize}

\subsection{Collection of numerical functions}
\label{app-collection_num_functions}

We collect here all the numerical functions obtained from the regularization
 procedure applied in Appendix \ref{sect-app-expansion_term_by_term}:
\beq
\mathcal{I}_{\mathcal{B}}^{(0)} (\gamma) \equiv \int_0^{\mu_0} d \mu \, \left\lbrace   
f(\mu)^{3/2} \left[  \le \sin \le \mu_0 - \mu \ri +1 \ri \log \le  \sin \le \mu_0 - \mu \ri +1 \ri   \right]
- \frac{1}{\le \mu_0 - \mu \ri^2} - \frac{1}{2 \le \mu_0 - \mu \ri}
\right\rbrace  \, ,
\label{eq:numerical_function_IB0}
\eeq
\beq
\mathcal{I}_{\mathcal{B}}^{(1)} (\gamma) \equiv \int_0^{\mu_0} d \mu \, \left[
f(\mu)^{3/2} \sin \le \mu_0 - \mu \ri 
- \frac{1}{\le \mu_0 - \mu \ri^2}
\right]  \, ,
\label{eq:numerical_function_IB1}
\eeq
\beq
\mathcal{I}_{\rm GHY} (\gamma) \equiv \int_0^{\mu_0}d\mu \sqrt{f(\mu)}\sin(\mu_0-\mu) \, ,
\label{eq:numerical_function_IGHY}
\eeq
\beq
\mathcal{I}_{\mathcal{J}}^{(0)} (\gamma) \equiv \int_0^{\mu_0}d\mu \sqrt{f(\mu)}\sin(\mu_0-\mu)\log\left( \sqrt{f(\mu)} \sin(\mu_0-\mu) \right) \; ,
\label{eq:numerical_function_J0}
\eeq
\beq
\mathcal{I}_{\mathcal{J}}^{(1)} (\gamma) \equiv 
 \int_0^{\mu_0} d\mu \, \left\lbrace \sqrt{f(\mu)} \, \log \left[   \sqrt{f(\mu)} \le 1+ \sin \le \mu_0 - \mu \ri \ri \right] 
+  \frac{1}{\mu_0 - \mu} \log \le \mu_0 - \mu \ri  \right\rbrace \, ,
\label{eq:numerical_function_J1}
\eeq
\beq
\begin{aligned}
\mathcal{I}_{\rm ct}^{(0)} (\gamma) \equiv
\int_0^{\mu_0} d\mu \, &
\left\lbrace - 2 \sqrt{f(\mu)} \le 1- \sin \le \mu_0 - \mu \ri \ri 
\log \left[ \le \sin \le \mu_0 - \mu \ri + \frac{f'(\mu)}{2 f(\mu)} \cos \le \mu_0 - \mu \ri \ri \right. \right. \\
& \left. \left. \times  \le 1+ \sin \le \mu_0 - \mu \ri \ri \right]  - \frac{2}{\mu_0 - \mu} \log \le \mu_0 - \mu \ri  \right\rbrace \, ,
\end{aligned}
\label{eq:numerical_function_Ict0}
\eeq
\beq
\mathcal{I}_{\rm ct}^{(1)} (\gamma)  \equiv
\int_0^{\mu_0} d \mu \, \left[ \frac{-2 \sqrt{f(\mu)} K(\mu)}{2 f(\mu) \sin \le \mu_0 - \mu \ri + f'(\mu) \cos \le \mu_0 - \mu \ri}  
- \frac{2}{\le \mu_0 - \mu \ri^2} \right]
\, ,
\label{eq:numerical_function_Ict1}
\eeq
\beq
\begin{aligned}
K(\mu) \equiv  & \,
2 \cos^2 \le \mu_0 - \mu \ri f(\mu) + \cos^2 \le \mu_0 - \mu \ri \frac{f'(\mu)^2}{f(\mu)}  \\
& - \sin \le \mu_0 - \mu \ri \cos \le \mu_0 - \mu \ri f'(\mu) - \cos^2 \le \mu_0 - \mu \ri  f''(\mu) \, .
\end{aligned}
\eeq


\section{Counterterms on timelike boundaries}
\label{app-timelike_counterterm}

In this Appendix we consider the inclusion in the action of the counterterm
introduced in  \cite{Akhavan:2019zax,Omidi:2020oit}, which for $d=2$ is
\beq
I_{\rm ct}^{\rm cutoff} = - \frac{1}{8 \pi G} \int d x \, dt \, \sqrt{-h} \, 
\frac{1}{L}   \, ,
\label{eq:timelike_counterterm}
\eeq
where $h$ is the metric determinant of the induced metric on the boundary.
This term was  introduced in \cite{Akhavan:2019zax, Omidi:2020oit}
 for the regularization prescription A in Fig.~\ref{fig-2regs},
 in order to reproduce the divergences of regularization B.

\subsection{Janus AdS$_3$ geometry}

In the Janus AdS$_3$ background there are two timelike regulator surfaces:
\begin{itemize}
\item  The first timelike cutoff surface corresponds to $z=\delta,$ and its contribution reads
\beq
I^{\delta}_{\rm ct} = - \frac{L}{2 \pi G} \int_0^{\mu^*(\delta)} d\mu \int_{\delta}^{t_{\rm WDW} (\delta, \mu)} dt \,
\frac{f(\mu)}{\delta} = 
 - \frac{L}{2 \pi G} \int_0^{\mu^*(\varepsilon)} d\mu \, f(\mu) \, \sin \le \mu_0 - \mu \ri  \, ,
\eeq
where we put a symmetry factor of 4.
The integrand gives rise to divergences due to the singularity in $\mu=\mu_0.$
Therefore we Laurent-expand the function around this point
\beq
f(\mu) \sin \le \mu_0 - \mu \ri = \frac{1}{\mu_0 - \mu} + \mathcal{O} \le \mu_0 - \mu \ri \, .
\eeq
Adding and subtracting this divergence allows to find
\beq
I_{\rm ct}^{\delta} = -\frac{L}{2 \pi G} \left[ \log \le \frac{\mu_0}{\varepsilon} \ri + \mathcal{I}_{\rm ct}^{\delta} (\gamma)  \right] + \mathcal{O} (\varepsilon) \, ,
\eeq
where we define the numerical function
\beq
\mathcal{I}_{\rm ct}^{\delta} (\gamma) \equiv \int_0^{\mu_0} d\mu \, 
\le f(\mu) \sin \le \mu_0 - \mu \ri - \frac{1}{\mu_0 - \mu} \ri \, ,
\eeq
such that $\mathcal{I}_{\rm ct}^{\delta} (0)= \log \le \frac{4}{\pi} \ri .$
After subtracting the vacuum AdS solution, we find
\beq
\Delta I_{\rm ct}^{\delta} = - \frac{L}{2 \pi G} \left[\log \le \frac{\mu_0}{2} \ri + \mathcal{I}_{\rm ct}^{\delta} (\gamma)   \right] \, .
\label{eq:timelike_ct1}
\eeq
\item
The second timelike cutoff surface corresponds to $\mu=\mu^*(\varepsilon)$ and to its partner located at $\mu=-\mu^*(\varepsilon),$ which contributes to the same result by symmetry reasonings.
We find
\beq
\begin{aligned}
 I^{\varepsilon}_{\rm ct} & = - \frac{L}{2 \pi G \varepsilon^2} 
\left\lbrace \int_{\delta}^{z_{\rm int} (\mu^*(\varepsilon))} \frac{dz}{z} \, \sin \le \mu_0 - \mu^*(\varepsilon) \ri  
+ \int_{z_{\rm int}(\mu^*(\varepsilon))}^{z_{\rm RT}} \frac{dz}{z^2} \, \le \frac{l}{2} - z \ri  \right\rbrace  = \\
& = \frac{L}{2 \pi G} \left[ \frac{1}{\varepsilon} \log \le \frac{2 \delta}{l} \ri - \frac{1}{2}  \right] + \mathcal{O} (\varepsilon) \, .
\end{aligned}
\eeq
After subtracting the vacuum AdS solution, this simply vanishes, 
$ \Delta I_{\rm ct}^{\varepsilon} = 0$.
\end{itemize}
Therefore, the total contribution coming from timelike counterterms in the Janus AdS background amounts to a finite part given in eq.~\eqref{eq:timelike_ct1}.

\subsection{AdS$_3$/BCFT$_2$ model}

The AdS/BCFT model contains a cutoff surface located at $z=\delta .$ 
Since the extrinsic curvature on the cutoff surface at $z=\delta$ is $K=2/L,$ the timelike counterterm \eqref{eq:timelike_counterterm} is proportional to the corresponding GHY contribution:
\beq
I_{\rm ct}^{\delta} = - \frac{1}{2} I^{\delta}_{\rm GHY} = 
- \frac{L}{8 \pi G} \left[ \frac{\cot \alpha}{\sin\alpha} - \log \le \tan \le \frac{\alpha}{2} \ri \ri  \right] \, .
\eeq
The vacuum solution corresponds to $\alpha= \frac{\pi}{2},$ but for this value the action vanishes.
Therefore we directly obtain that the same result holds after the subtraction, $\Delta I_{\rm ct}^{\delta} = I_{\rm ct}^{\delta}$.


\bibliography{at}
\bibliographystyle{at}

\end{document}